\documentclass[lettersize,journal]{IEEEtran}
\usepackage{amsmath,amsfonts}
\usepackage{algorithmic}
\usepackage{array}
\usepackage[caption=false,font=normalsize,labelfont=sf,textfont=sf]{subfig}
\usepackage{textcomp}
\usepackage{stfloats}
\usepackage{url}
\usepackage{verbatim}
\usepackage{graphicx}
\def\BibTeX{{\rm B\kern-.05em{\sc i\kern-.025em b}\kern-.08em
    T\kern-.1667em\lower.7ex\hbox{E}\kern-.125emX}}
\usepackage{balance}

\usepackage{mathrsfs}
\usepackage{amsmath}
\usepackage{amsfonts}
\usepackage{amsthm}
\usepackage{amssymb}
\usepackage{graphicx}
\usepackage{indentfirst} 
\usepackage{array}
\usepackage{epstopdf}
\usepackage{cite}
\usepackage{enumerate}
\usepackage{bm}
\usepackage{dsfont}
\usepackage{multirow}
\usepackage{verbatim}
\usepackage{stfloats}
\usepackage{makecell}
\usepackage{cancel}
\usepackage{threeparttable}
\usepackage{tikz}
\usetikzlibrary{positioning}
\usetikzlibrary{spy}
\usepackage{wrapfig}
\usepackage{pgfplots}
\usepackage{nicefrac}
\usepackage{thmtools,thm-restate}
\usepackage{color}

\usepackage{booktabs}
\usepackage{bbm}
\usepackage{adjustbox}
\usepackage{multirow}
\usepackage{tabularx}
\usepackage[misc]{ifsym}
\usepackage{bbding}
\usepackage{colortbl}
\usepackage{algorithm}
\usepackage{xcolor}
\usepackage{hyperref}

\definecolor{lightblue}{rgb}{0.9, 0.92, 0.96}
\definecolor{subsectioncolor}{rgb}{0, 0, 0.706}   
\definecolor{mylightblue}{rgb}{0.851, 0.898, 0.941} %
\definecolor{LightCyan}{rgb}{0.88,1,1}
\definecolor{LightOrange}{rgb}{1,0.85,0.70}

\usepackage{pifont}
\newcommand{\cmark}{\ding{51}}%
\newcommand{\xmark}{\ding{55}}%

\definecolor{C0}{rgb}{0.121569, 0.466667, 0.705882}
\definecolor{C1}{rgb}{1.000000, 0.498039, 0.054902}
\definecolor{C2}{rgb}{0.172549, 0.627451, 0.172549}
\definecolor{C3}{rgb}{0.839216, 0.152941, 0.156863}
\definecolor{C4}{rgb}{0.580392, 0.403922, 0.741176}
\definecolor{C5}{rgb}{0.549020, 0.337255, 0.294118}
\definecolor{C6}{rgb}{0.890196, 0.466667, 0.760784}
\definecolor{C7}{rgb}{0.498039, 0.498039, 0.498039}
\definecolor{C8}{rgb}{0.737255, 0.741176, 0.133333}
\definecolor{C9}{rgb}{0.090196, 0.745098, 0.811765}
\definecolor{trolleygrey}{rgb}{0.5, 0.5, 0.5}

\begin{document}
\title{Error-Resilient Semantic Communication for Speech Transmission over Packet-Loss Networks}
\author{
Zhuohang Han,
Jincheng Dai,~\IEEEmembership{Member, IEEE},
Shengshi Yao,~\IEEEmembership{Member, IEEE},
Junyi Wang,~\IEEEmembership{Member, IEEE},
Yanlong Li,~\IEEEmembership{Member, IEEE},
Kai Niu,~\IEEEmembership{Member, IEEE},
Wenjun Xu,~\IEEEmembership{Senior Member, IEEE},
and Ping Zhang,~\IEEEmembership{Fellow, IEEE}
\thanks{This work was supported in part by the National Key Research and Development Program of China under Grant 2024YFF0509700; in part by the National Natural Science Foundation of China under Grant 62371063, Grant 62321001, and Grant 62501075; in part by the Beijing Municipal Natural Science Foundation under Grant L232047; and in part by the Beijing Nova Program.}

\thanks{Zhuohang Han, Jincheng Dai, Shengshi Yao, Kai Niu, Wenjun Xu, and Ping Zhang are with Beijing University of Posts and Telecommunications, Beijing, China. Junyi Wang and Yanlong Li are with the Academy of Information and Communication, Guilin University of Electronic Technology, Guangxi, China.}

}


\maketitle

\begin{abstract}
Real-time speech communication over wireless networks remains challenging, as conventional channel protection mechanisms cannot effectively counter packet loss under stringent bandwidth and latency constraints. 
Semantic communication has emerged as a promising paradigm for enhancing the robustness of speech transmission by means of joint source-channel coding (JSCC). 
However, its cross-layer design hinders practical deployment due to the incompatibility with existing digital communication systems.
In this case, the robustness of speech communication is consequently evaluated primarily by the error-resilience to packet loss over wireless networks. 
To address these challenges, we propose \emph{Glaris}, a generative latent-prior-based resilient speech semantic communication framework that performs resilient speech coding in the generative latent space.
Generative latent priors enable high-quality packet loss concealment (PLC) at the receiver side, well-balancing semantic consistency and reconstruction fidelity.
Additionally, an integrated error resilience mechanism is designed to mitigate the error propagation and improve the effectiveness of PLC.
Compared with traditional packet-level forward error correction (FEC) strategies, our new method achieves enhanced robustness over dynamic wireless networks while reducing redundancy overhead significantly.
Experimental results on the LibriSpeech dataset demonstrate that \emph{Glaris} consistently outperforms existing error-resilient codecs, achieving JSCC-level robustness while maintaining seamless compatibility with existing systems, and it also strikes a favorable balance between transmission efficiency and speech reconstruction quality.
\end{abstract}

\begin{IEEEkeywords}
Semantic communication, neural speech coding, packet loss concealment, forward error correction, efficient redundancy.
\end{IEEEkeywords}

\section{Introduction}\label{section_introduction}
\IEEEPARstart{R}{eal-time} speech communication has become a cornerstone of modern digital services, including cloud-native 4G/5G voice calls, online meetings, cloud-gaming voice chat, and voice-enabled edge applications. 
These latency-sensitive scenarios demand stringent end-to-end delay guarantees and robustness against packet loss, as the quality of experience is highly sensitive to latency, jitter, and loss~\cite{TIP2024QoE}. 
Retransmission-based recovery, though effective in non-real-time data applications, is infeasible for interactive speech, because the round-trip delay of automatic repeat request mechanisms typically exceeds acceptable conversational latency budget~\cite{perkins1998survey}. 
Hence, achieving reliability within one-shot transmission remains a key challenge for real-time speech communication systems.

Semantic communication has recently emerged as a promising paradigm to address this challenge by enabling more robust communication under unreliable channels~\cite{gunduz2022beyond}. Semantic communication systems often use joint source-channel coding (JSCC) to extract and transmit high-level semantic representations across modalities such as text~\cite{xie2021deep}, images~\cite{bourtsoulatze2019deep,dai2022nonlinear,zhang2023predictive}, speech~\cite{weng2021semantic}, and video~\cite{wang2022wireless}. However, this cross-layer design makes them incompatible with standard protocol stacks and existing modulation and coding schemes, limiting deployability. Additionally, JSCC is typically trained for specific channel conditions, which hinders its adaptation to dynamic wireless environments without compromising performance.
To address these challenges, we revisit semantic communication as a joint optimization problem of source compression and network transmission, aiming to achieve JSCC-level robustness over packet-loss networks using traditional physical-layer transmission. This is nontrivial, as packet loss removes entire packets rather than corrupting individual symbols, resulting in complete information loss and making error recovery significantly more difficult than physical-layer impairments.

To address this problem, we draw inspiration from traditional error-resilience mechanisms above the network layer, which can be broadly categorized into sender-based forward error correction (FEC) and receiver-based packet loss concealment (PLC). 
Sender-based FEC introduces controlled redundancy to enable packet recovery without retransmission, as in out-of-band FEC (e.g., Reed-Solomon, fountain codes~\cite{wicker1999reed,mackay2005fountain}), redundant encoding~\cite{RFC2198}, and codec-specific in-band redundancy (e.g., Opus LBRR~\cite{valin2012definition}).
RFC8854~\cite{RFC8854} recommends the use of in-band FEC when available, and out-of-band FEC is therefore beyond the scope of this work. Nevertheless, in-band FEC like LBRR provides only single-frame protection and is ineffective against burst or consecutive losses. Hence, receiver-based PLC is often employed as a complementary approach, which can be categorized as heuristic or neural.
Traditional PLC relies on waveform repetition or spectral interpolation~\cite{perkins1998survey}, while neural PLC employs deep generative models~\cite{shi2019speech,pascual2021adversarial,wang2021temporal,li2022end,valin2024very} to reconstruct lost segments. 
Despite recent progress in neural PLCs, two main limitations remain: (i) each frame must be independently decodable to provide context for loss prediction, which prevents the use of entropy coding, thereby limiting compression efficiency; and (ii) limited inter-frame correlation restricts the achievable reconstruction performance under severe or burst packet loss.

In this paper, we propose a standard-compatible semantic communication framework that integrates sender-based in-band FEC and receiver-based PLC into codec design, aimed at ensuring robust speech transmission over packet-loss networks. Our goal is to improve reconstruction quality under severe or burst losses for PLC, while reducing redundancy overhead and increasing robustness to dynamic channel conditions for in-band FEC. However, designing such a framework presents two primary challenges.

The first challenge is maintaining semantic consistency in concealed speech under packet loss for PLC. Existing approaches~\cite{shi2019speech,pascual2021adversarial,wang2021temporal,li2022end,valin2024very} primarily focus on fine-grained acoustic details but fail to model long-range dependencies in a single-stage framework. Due to strong local correlations in speech, these methods tend to over-rely on short-term cues, which results in degraded phoneme and word-level coherence. To address this issue, we introduce generative latent priors as a regularization mechanism to guide the learning of long-range dependencies, thereby enhancing perceptual naturalness.

The second challenge lies in balancing compression efficiency with error resilience. It requires efficient in-band FEC design and methods to suppress error propagation in entropy-coded streams, enabling the use of entropy coding for high-efficiency compression under packet-loss conditions. While compression minimizes redundancy to improve efficiency, error resilience depends on redundancy for recovery, resulting in an intrinsic trade-off between compression efficiency and robustness. This trade-off is pronounced in entropy-coded streams, where symbol dependencies can cause cascading errors. To address this, we reuse hyperprior-derived side information as in-band FEC to provide redundancy for both entropy decoding and PLC, thereby suppressing error propagation and improving reconstruction fidelity under packet loss.

Recent advances in image compression~\cite{jia2024generative, qi2025generative} have adapted a tokenization and compression architecture and have achieved remarkable rate-distortion (RD) results. Inspired by this, we propose \emph{Glaris}, a \textbf{G}enerative \textbf{La}tent-prior-based \textbf{R}es\textbf{i}lient \textbf{S}peech semantic communication framework, which leverages generative latent priors within a two-stage coding architecture to achieve error-resilient speech transmission. In the first stage, a VQ-VAE preserves fine-grained acoustic details, enabling faithful reconstruction. In the second stage, error-resilient transform coding with latent-prior modeling captures long-range dependencies, enhancing both semantic consistency and reconstruction fidelity. The hyperprior serves as a compact and effective redundancy, functioning as in-band FEC to: (i) guide PLC via high-level contextual cues, (ii) suppress entropy decoding errors, and (iii) minimize bitrate overhead by encoding latent distributions rather than raw features. These generative latent priors, encompassing both latent-space priors and hyperpriors, form the foundation of \emph{Glaris}, enabling improved semantic consistency and a favorable balance between compression efficiency and error resilience under burst losses.

We evaluate \emph{Glaris} on the LibriSpeech dataset across diverse packet-loss conditions, including independent and identically distributed (i.i.d.) random losses, three-state Markov channels~\cite{milner2004analysis}, actual traces from PLC challenge dataset~\cite{diener2022interspeech}, and COST2100 wireless channels.
Experimental results demonstrate that \emph{Glaris} achieves strong robustness against high loss rates and long bursts, achieving a favorable balance between compression efficiency and resilience compared with both separation-based and JSCC-based baselines.
Subjective listening tests further confirm the perceptual benefits and practical applicability of the proposed framework. 
Moreover, real-time factor (RTF) evaluations demonstrate that \emph{Glaris} supports real-time streaming inference, making it suitable for deployment in real-world speech communication systems.

Our key contributions are summarized as follows:
\begin{enumerate}
    \item \emph{Standard-Compatible Semantic Communication Framework:}
	We propose \emph{Glaris}, an error-resilient semantic communication framework for speech transmission over packet-loss networks, which performs error-resilient transform coding in the generative latent space of a VQ-VAE to achieve semantic consistency and high reconstruction fidelity under packet loss.

	\item \emph{Side-Information-Based Error-Resilience Enhancement:}
    We design an error resilience mechanism that incorporates side-information-based in-band FEC into PLC design to effectively suppress error propagation and guide accurate prediction of lost frames.
  
	\item \emph{Controllable Redundancy:}
    \emph{Glaris} enables adaptive redundancy control through side information rate and backup frame configuration, offering efficient robustness adjustment under varying channel conditions.
\end{enumerate}

The remainder of this paper is organized as follows. 
Section~\ref{sec:relted work} reviews related studies on neural speech coding and error resilience mechanisms. 
Section~\ref{section_method} introduces the proposed framework, 
Section~\ref{sec:exp} presents the experimental evaluations, 
and Section~\ref{sec:con} concludes the paper.

\section{Related Work}\label{sec:relted work}

\subsection{Neural Audio/Speech Coding}  
Neural audio/speech coding can typically be divided into neural vocoders and end-to-end neural coding.
Neural vocoders based on WaveNet~\cite{kleijn2018wavenet} apply neural networks as the decoder to decode from traditional handcrafted features like spectral envelope, pitch, and voicing level. LPCNet~\cite{valin2019lpcnet} further improves efficiency by combining neural modeling with linear prediction, achieving real-time speech coding at 1.6 kbps. 
To leverage the full potential of neural coding, end-to-end neural coding has been introduced to obtain learned features. Based on the SENet structure~\cite{roblek2020seanet} and VQ-VAE framework~\cite{van2017neural}, a series of works~\cite{zeghidour2021soundstream, defossez2023high, wu2023audiodec, yang2023hifi, kumar2023high} have been proposed. SoundStream~\cite{zeghidour2021soundstream} trains the model with residual vector quantization (RVQ) in an adversarial learning strategy to improve the perceptual quality. Based on that, Encodec~\cite{defossez2023high} employs an RNN to improve the sequence modeling and trains a small transformer to predict the distribution of codewords to further improve the compression efficiency.
Instead of using entropy coding to improve compression efficiency, HiFi-Codec~\cite{yang2023hifi} and Descript-Audio-Codec (DAC)~\cite{kumar2023high} 
design the new structure of RVQ with a well-designed training strategy to improve the utilization of the codebook. 

However, advancements in the field of learned image compression~\cite{balle2017end, balle2018variational,jia2024generative} 
face the RD optimization using variational methods with scalar quantization (SQ), which is rarely explored in neural speech codecs. 
Several neural speech codecs based on SQ have been proposed, but they operate at comparatively high bitrates~\cite{shin2022deep,byun2023perceptual}. An SQ-based neural speech coding scheme for transmitting redundant information in order to increase robustness against transmission errors has been proposed by~\cite{valin2024dred}. ~\cite{brendel2024neural} applies the finite SQ in a speech codec to ensure constant packet lengths. 
However, none of the existing neural speech codecs have adopted the tokenization and compression architecture proposed in~\cite{jia2024generative}, which motivates us to introduce this structure in speech coding to achieve efficient compression and enhanced error resilience through dedicated mechanism design.

\subsection{Error Resilience Mechanism}
Error resilience mechanisms can be divided into sender-based and receiver-based approaches~\cite{perkins1998survey}. Sender-based mechanisms mainly include retransmission and packet-level FEC~\cite{wicker1999reed,mackay2005fountain}, while retransmission introduces unacceptable delay for VoIP and packet-level FEC is difficult to adapt to dynamic channel states when optimized for efficiency. In this case, receiver-based mechanism PLC plays a more important role in VoIP. Traditional PLC methods include zero filling, interpolation, and comfortable background noise~\cite{perkins1998survey}. Nowadays, interpolation using a neural network called neural PLC has achieved great success, especially for GAN-based PLC~\cite{shi2019speech,pascual2021adversarial,wang2021temporal,li2022end,valin2024very}. However, as described in FD-PLC~\cite{xue2022towards}, post-processed PLC is inherently constrained by the decoder. DRED~\cite{valin2024dred} follows a similar paradigm, since it introduces a neural encoder to encode features of Opus into the low-bitrate deep redundancy bitstream, and invokes neural PLC in features when redundancy is unavailable. Similar to DRED, further works~\cite{yao2025soundspring, kolundvzija2024low,gupta2024improving} design deep redundancy schemes based on discrete RVQ tokens for neural codecs that perform well in the low-bitrate range. In contrast, our work leverages the inherent hyperprior of the codec as deep redundancy, without introducing any additional encoder or decoder, and demonstrates its effectiveness in improving error resilience.

\section{Methodology}\label{section_method}
\begin{figure*}[t]
	\setlength{\abovecaptionskip}{0cm}
	\setlength{\belowcaptionskip}{0.cm}
	\hspace{-1em}
	\centering
	\includegraphics[width=0.9\linewidth]{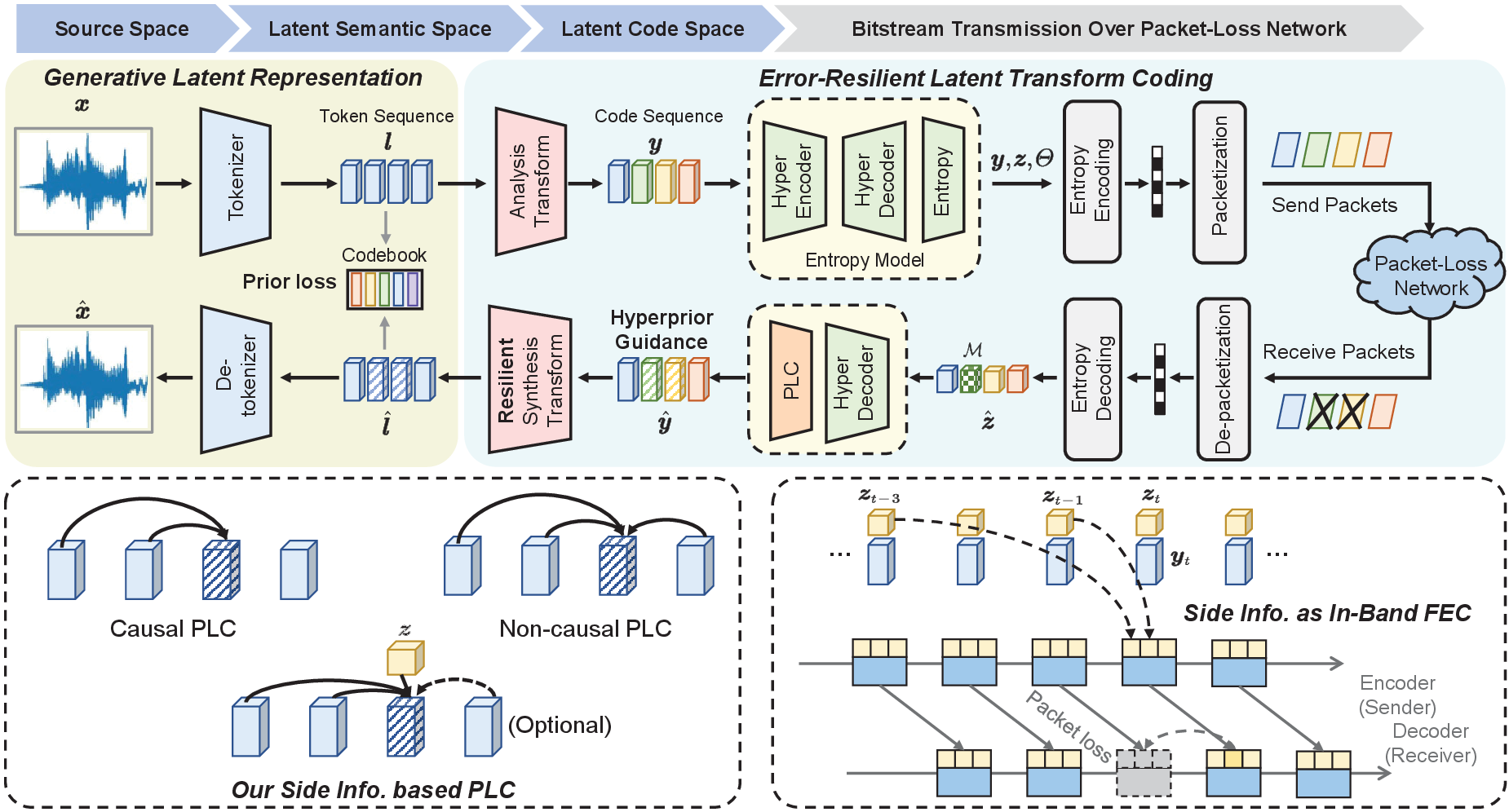}
	\caption{
	Proposed error-resilient speech communication framework using generative latent priors. 
    Top: Overview of the two-stage coding framework, where a VQ-VAE first learns a generative latent representation that preserves fine-grained acoustic features, and the subsequent transform coding stage operates on the generative latent space for RD optimization and error resilience. Latent prior loss and hyperprior guidance further enhance semantic consistency for compression and error resilience. 
    Bottom left: Proposed side-information-based PLC. 
    Bottom right: Side information reused as in-band FEC.
	}
	\label{fig:framework}
\end{figure*}
\subsection{Overview} 
\emph{Glaris} enhances the error resilience of speech communication systems by using generative latent priors within a two-stage coding framework. As illustrated in the top part of Fig.~\ref{fig:framework}, a VQ-VAE first learns a generative latent representation that captures high-level semantic information, enabling high-fidelity reconstruction. The subsequent error-resilient transform coding stage then operates on this latent space to achieve RD optimization and enhance robustness under packet loss conditions. The latent prior is introduced as a regularization loss to enforce sequence-level consistency, whereas the hyperprior provides high-level guidance for PLC. Through the joint use of prior loss and hyperprior guidance, \emph{Glaris} achieves an effective balance between compression efficiency and robustness under unreliable transmission.


The overall data processing flow is summarized as follows. The input speech signal $\bm{x}$ is encoded into a generative latent representation $\bm{l} = E(\bm{x})$ by latent encoder $E(\cdot)$. Next the token sequence $\bm{l}$ is transformed into latent code $\bm{y} = g_a(\bm{l})$ by the analysis transform $g_a(\cdot)$. The latent code $\bm{y}$ is scalar-quantized to $\bm{y}_Q = Q(\bm{y})$, and its quantized symbols are entropy coded for transmission over a lossy channel $W(\cdot)$ along with the side information $\bm{z}$ derived from the hyperprior model. At the receiver side, the received latent code $\hat{\bm{y}}$ is decoded by the synthesis transform $g_s(\cdot)$ to reconstruct $\hat{\bm{l}}$, and the latent decoder $D(\cdot)$ generates the reconstructed speech $\hat{\bm{x}}$. The side information $\bm{z}$ not only captures the distribution of $\bm{y}$ but also provides auxiliary cues for PLC, thereby improving both efficiency and robustness. The procedure of \emph{Glaris} can be summarized as
\begin{equation}
\bm{x}
\xrightarrow{E(\cdot)}
\bm{l}
\xrightarrow{g_a(\cdot)}
\bm{y}
\xrightarrow{Q(\cdot)}
\bm{y}_Q
\xrightarrow{W(\cdot)}
\hat{\bm{y}}
\xrightarrow{g_s(\cdot)}
\hat{\bm{l}}
\xrightarrow{D(\cdot)}
\hat{\bm{x}}.
\label{eq:process}
\end{equation}

To further enhance error resilience, a side-information-based error resilience mechanism is designed as illustrated in the bottom part of Fig.~\ref{fig:framework}. The key insight is that, for a lost frame $\bm{y}_t$, the directly encoded latent $\bm{z}_t$ provides stronger correlation and richer reconstruction cues than context frames. To ensure reliable reception of $\bm{z}_t$, the side information is reused as in-band FEC, whose redundancy level can be adaptively controlled through the bitrate of $\bm{z}$ and backup frame configuration. By introducing an offset in backups, \emph{Glaris} achieves robust recovery against long burst losses while maintaining compression efficiency.

\subsection{Two-Stage Coding}\label{sec:Latent Transform Coding}

\begin{figure}[t]
	\setlength{\abovecaptionskip}{0cm}
	\setlength{\belowcaptionskip}{0.cm}
	\hspace{-1em}
	\centering
	\includegraphics[width=0.95\linewidth]{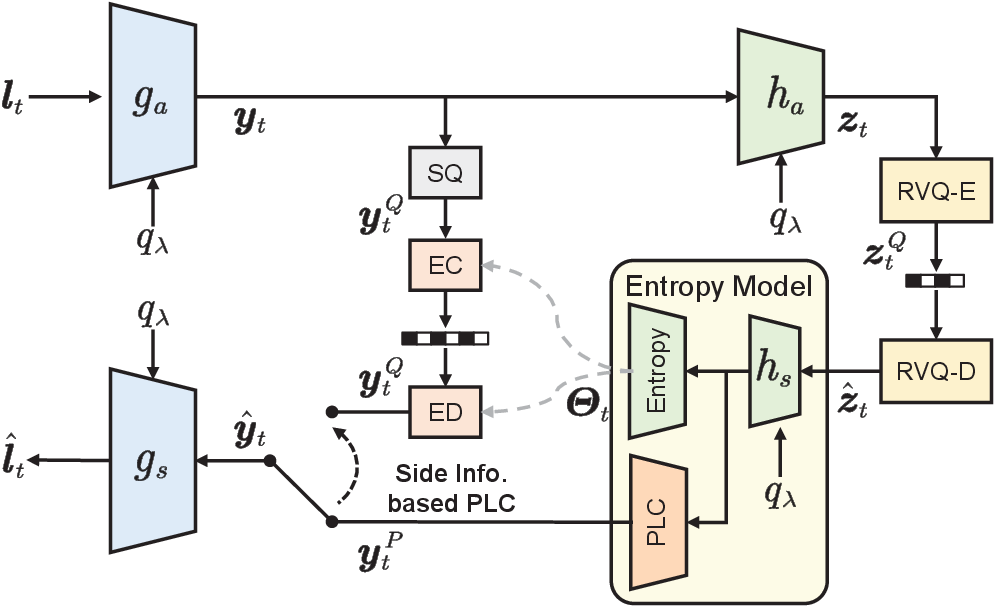}
	
	\caption{Illustration of the proposed error-resilient latent transform coding with side-information-based PLC, where the hyperprior is compressed by RVQ.
    A dual-function entropy model utilizes the decoded side information for both entropy decoding and PLC. 
    In error-free transmission, the entropy module predicts the distribution parameters for entropy decoding, whereas under packet loss, the PLC module reconstructs the missing latent frame using the received side information.
    }
	\label{fig:ntc}
\end{figure}

\subsubsection{Generative Latent Representation}
A key challenge in learning a high-quality generative latent representation lies in constructing a manifold-aligned latent space that preserves the acoustic features of speech. 
In Glaris, this is achieved by employing a VQ-VAE~\cite{zeghidour2021soundstream} as the latent audio encoder-decoder pair. 
The VQ-VAE encodes high-dimensional speech representations into a compact latent space, reducing dimensionality while preserving semantic structure and perceptual consistency.
This compact space enables tractable latent priors that are leveraged during training to enforce global sequence consistency. 
The discrete codebook acts as a variational bottleneck, enforcing compact and robust latent representations, thereby enhancing both compression efficiency and error resilience.

\subsubsection{Error-Resilient Latent Transform Coding}
A straightforward way to compress the token sequence $\bm{l}$ is the VQ-indices-map coding~\cite{zeghidour2021soundstream}. 
Although EnCodec~\cite{defossez2023high} improves compression efficiency by learning the distribution of discrete indices, transform coding enables explicit RD optimization through variational modeling, thereby achieving higher compression efficiency, as shown in Fig.~\ref{fig:RD_PESQ_0}. 
Building upon this principle, \emph{Glaris} introduces an error-resilient latent transform coding framework to enhance the error resilience while maintaining compression efficiency.

The proposed architecture, illustrated in Fig.~\ref{fig:ntc}, employs a dual-function entropy model that supports both entropy coding and PLC. 
For each latent frame $\bm{l}_t$, the analysis transform $g_a(\cdot)$ produces $\bm{y}_t$, which is quantized to $\bm{y}_t^Q$. 
A hyper transform $h_a(\cdot)$ generates side information $\bm{z}_t$, later quantized using RVQ and transmitted along with $\bm{y}_t^Q$. 
At the receiver, the shared hyper synthesis transform $h_s(\cdot)$ decodes $\hat{\bm{z}}_t$, which is used by two subsequent modules:
\begin{equation}
\bm{\varTheta}_t = f_{\text{entropy}}(h_s(\hat{\bm{z}}_t)), \quad 
\bm{y}_t^{P} = f_{\text{PLC}}(h_s(\hat{\bm{z}}_t)),
\end{equation}
where $f_{\text{entropy}}(\cdot)$ and $f_{\text{PLC}}(\cdot)$ represent entropy and PLC module respectively.
When no packet loss occurs, Gaussian distribution parameters $\bm{\varTheta}_t$ are used for entropy coding. 
When frame $t$ is lost, $\bm{y}_t^{P}$ is used as a substitution for $\hat{\bm{y}}_t$, providing hyperprior for latent recovery. 
This dual-function design allows the side information to provide a hyperprior for PLC, significantly improving robustness under packet loss.

To further enhance resilience, the synthesis transform $g_s(\cdot)$ is trained with simulated loss patterns, enabling the model to reconstruct missing or corrupted latents with contextual awareness. 
Thus, robustness is inherently built into the decoder through joint optimization under loss-perturbed conditions.

All neural modules are implemented using a causal streaming transformer, which improves long-range sequence modeling while supporting streaming transmission. 
The transform modules are conditioned on rate control parameters $q_\lambda$, where $q_\lambda$ represents quantized controls of rate-related hyperparameters $\lambda$. 
This design allows \emph{Glaris} to achieve causal and contextual streaming inference while supporting flexible rate control.

\subsubsection{RVQ-based Hyperprior Module}
Most neural transform coding frameworks employ a factorized hyperprior~\cite{balle2018variational} to model the side information $\bm{z}$. Although effective for distribution learning, this design produces variable-length entropy-coded bitstreams for both $\bm{y}$ and $\bm{z}$, making their boundaries difficult to delimit within a single packet and thus complicating rate allocation for $\bm{z}$. 

To overcome these limitations, we introduce an RVQ-based hyperprior that replaces the infinite-codebook scalar quantization with a finite-codebook RVQ and indices-map coding, yielding a fixed-length, index-level representation of $\bm{z}$. By fixing $\bm{z}$ to a low bitrate, the side information is compelled to encode only the most informative features, effectively avoiding excessive bitrate overhead from $\bm{z}$. This not only enhances compression efficiency at low bitrates but also improves PLC performance with negligible additional bandwidth, as $\bm{z}$ provides highly informative cues for recovery. 

Although the proposed module supports multi-rate RVQ, a single low-rate setting for $\bm{z}$ is sufficient when only compression efficiency is considered, as the RD performance remains similar across $\bm{z}$ bitrates under a fixed resilience configuration (see Fig.~\ref{fig:RD_PESQ_0}). By contrast, increasing the bitrate of $\bm{z}$ directly benefits PLC, since richer side information yields more effective latent recovery under packet loss conditions. Hence, multi-rate control is introduced primarily to adapt redundancy for error resilience.

\begin{figure}[t]
	\setlength{\abovecaptionskip}{0cm}
	\setlength{\belowcaptionskip}{0.cm}
	\hspace{-1em}
	\centering
	\includegraphics[width=0.95\linewidth]{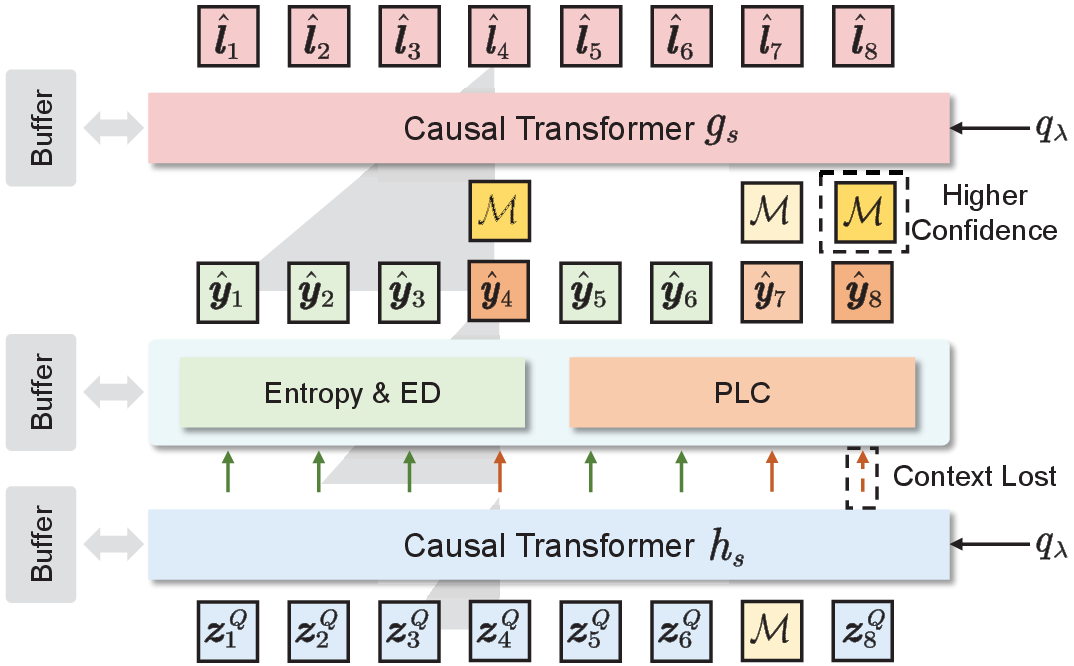}

	\caption{
	Illustration of the side-information-based PLC. 
    The decoding path is selected according to whether the current latent $\hat{\bm{y}}$ can be successfully entropy decoded, as reflected by the color of the arrows. 
    To alleviate context loss, the context length of both the entropy model and the transformer $h_s$ is restricted. 
    When the side information $\bm{z}$ is unavailable, a learned mask token serves as its substitute. 
    In the latent $\bm{y}$ space, another learned mask token is added to the predicted $\hat{\bm{y}}$ to represent the confidence of the reconstruction.
    Through this hierarchical process, the reconstructed latent representation $\hat{\bm{l}}$ integrates side information and inter-frame dependencies.
	}

	\label{fig:side_info_plc}
\end{figure}

\subsubsection{Rate-Variable Transformation}
In practical speech communication, bitrate control is crucial for adapting to dynamic channel bandwidth. Unlike RVQ, which adjusts bitrate through the number of quantizers, transform coding achieves finer and more flexible control by tuning the Gaussian parameters $\bm{\varTheta}$ in the entropy model. 

Rate control for transform coding is typically realized through quantization parameter tuning or hyper-parameter embedding~\cite{choi2019variable, li2022hybrid, li2024neural}. For simplicity, we adopt the latter, while noting that other rate control schemes can also be integrated into our framework. 

The RD objective is formulated as $\mathcal{L}_{\text{RD}} = \mathcal{D} + \lambda \mathcal{R}$, 
where $\mathcal{D}$ denotes distortion, $\mathcal{R}$ is the bitrate, and $\lambda$ controls their trade-off. 
We define a rate-control index $q_\lambda \in \{0, 1, \ldots, q_{\text{num}} - 1\}$ and compute $\lambda$ as:
\begin{equation}
    \lambda = \exp \left( \ln \lambda_{\text{min}} + 
    \frac{q_\lambda}{q_{\text{num}} - 1} (\ln \lambda_{\text{max}} - \ln \lambda_{\text{min}}) \right),
\end{equation}
where total quantization levels $q_{\text{num}}$ defaults to 64 during training. And $q_\lambda$ is uniformly sampled and embedded as a conditional vector to provide target bitrate information for the transform modules.





\subsection{Side-Information-based Error-Resiliency Enhancement}\label{sec:Side-Information-based Error-Resiliency Enhancement}

\begin{figure}[t]
	\setlength{\abovecaptionskip}{0cm}
	\setlength{\belowcaptionskip}{0.cm}
	\hspace{-1em}
	\centering
	\includegraphics[width=0.95\linewidth]{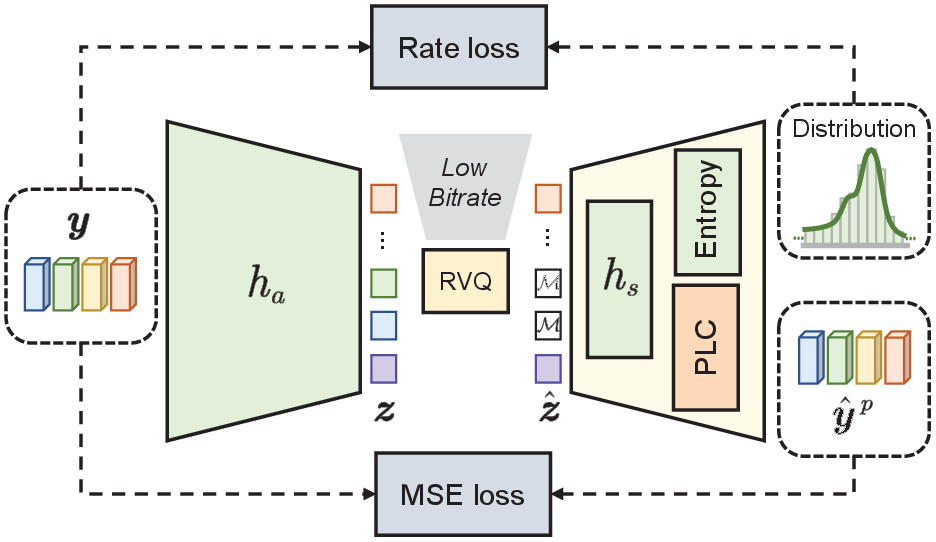}
	
	\caption{
	Learning process of the side information.
    The entropy module predicts the distribution of $\bm{y}$ for entropy coding, while the PLC module reconstructs $\bm{y}$ under MSE supervision.
    To constrain bitrate and enable rate control, the side information $\bm{z}$ is compressed with RVQ.
    During training, random masking with learned tokens $\mathcal{M}$ is applied to $\bm{z}$ at a ratio between 0 and 0.1 to improve robustness against missing side information.
	}
	\label{fig:side_info_learning}
\end{figure}
\subsubsection{Side-Information-based PLC}
Previous PLC methods predict missing frames from the surrounding context, but their performance is constrained by weak inter-frame correlation, especially under burst losses. This limitation becomes more severe in efficient compression systems where redundancy is minimized. To address this, we introduce side information $\bm{z}_t$ as an additional condition during inference:
\begin{equation}
\hat{\bm{y}}_{t} = f\!\left(\bm{y}_{t-n \leq \tau < t}, \bm{z}_t\right),
\end{equation}
where $n$ denotes the number of past frames used for causal prediction, and $\bm{z}_t$ provides informative cues for recovering lost content in streaming conditions.

As illustrated in Fig.~\ref{fig:side_info_plc}, the side-information-based PLC reuses the hyper synthesis transform $h_s(\cdot)$ to extract hyperprior shared by the entropy and PLC modules. During inference, $\bm{z}_t$ is first decoded through $h_s(\cdot)$ to produce loss-aware guidance. If $\bm{y}_t$ is successfully received, the output is utilized by the entropy module for decoding. Otherwise, the PLC module is activated to estimate $\hat{\bm{y}}_t$. When $\bm{z}_t$ is incomplete, the missing positions in $\bm{z}^Q$ are replaced with a learned mask token $\mathcal{M}$, forming a masked input $\hat{\bm{z}}$ that maintains reliable inference under side information loss. To reduce error propagation, the temporal context of $h_s$ and the entropy module is limited, while the PLC module accesses a longer context window to improve reconstruction fidelity and long sequence consistency under burst losses.

To represent prediction confidence, distinct mask tokens are applied depending on the availability of $\bm{z}_t$. High- and low-confidence tokens, denoted as $\mathcal{M}_H$ and $\mathcal{M}_L$, are incorporated into the PLC output to distinguish reliable and uncertain regions. The final reconstructed latent $\hat{\bm{y}}$ is computed as
\begin{equation}
\begin{aligned}
    \bm{y}_H^P &= f_{\text{PLC}}(h_s(\hat{\bm{z}})) + \mathcal{M}_H, \\
    \bm{y}_L^P &= f_{\text{PLC}}(h_s(\hat{\bm{z}})) + \mathcal{M}_L, \\
    \bm{y}_M^P &= \bm{y}_L^P \odot M_{\bm{z}} + \bm{y}_H^P \odot (1 - M_{\bm{z}}), \\
    \hat{\bm{y}} &= \bm{y}_M^P \odot M_{\bm{y}} + \bm{y}^Q \odot (1 - M_{\bm{y}}),
\end{aligned}
\end{equation}
where $M_{\bm{y}}$ and $M_{\bm{z}}$ are binary masks, in which a true value indicates missing positions.

This unified framework integrates side information and context-aware prediction within a causal streaming design, enabling robust recovery even under severe burst losses.

\subsubsection{In-band FEC via Side Information}

\begin{figure*}[t]
	\setlength{\abovecaptionskip}{0cm}
	\setlength{\belowcaptionskip}{0.cm}
	\hspace{-1em}
	\centering
	\includegraphics[width=0.95\linewidth]{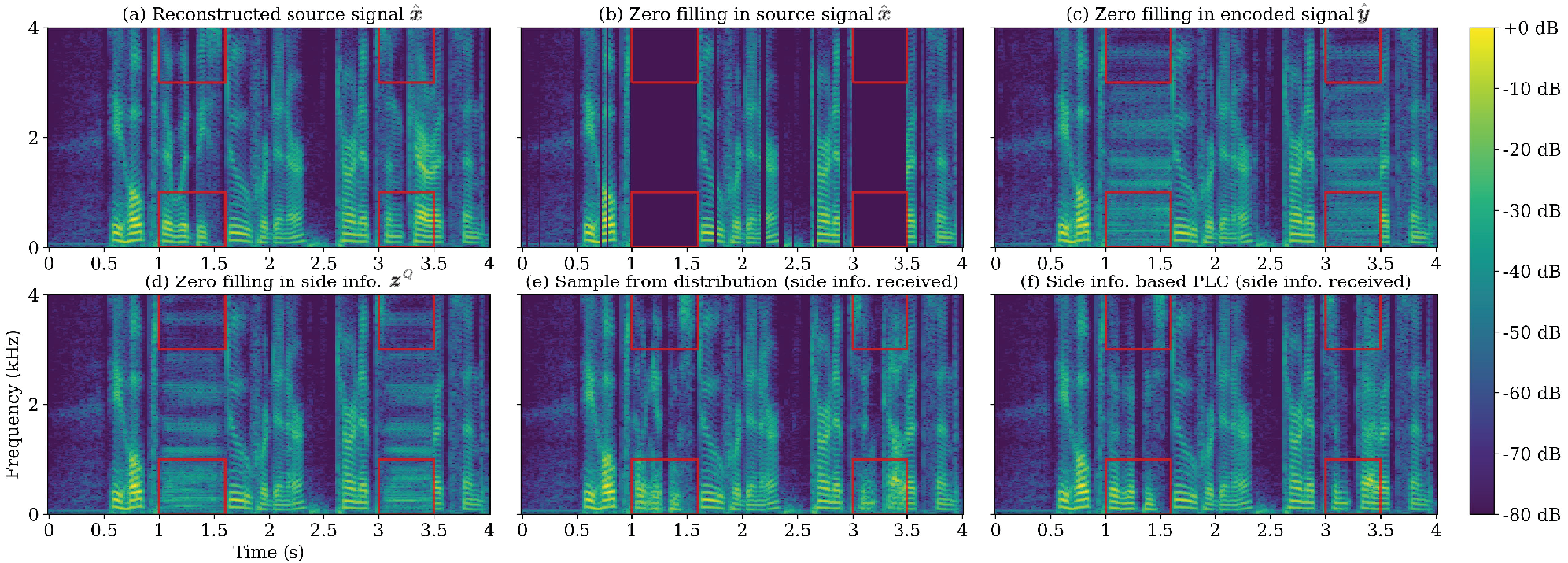}
	
	\caption{Magnitude spectrograms (in dB) of an example speech utterance under different PLC strategies. 
	(a) Reconstructed source signal without packet loss. (PESQ = 4.13, PLCMOS = 4.34, STOI = 0.99)
    (b) Zero filling in source signal $\hat{\bm{x}}$, where zero filling position are masked for visualization. (PESQ = 1.46, PLCMOS = 2.33, STOI = 0.82)
    (c) Zero filling in encoded signal $\hat{\bm{y}}$. (PESQ = 1.72, PLCMOS = 3.08, STOI = 0.87)
    (d) Zero filling in side information $\bm{z}^Q$. (PESQ = 1.88, PLCMOS = 3.05, STOI = 0.89)
    (e) Sampling from the predicted distribution. (PESQ = 2.30, PLCMOS = 3.69, STOI = 0.93)
    (f) Proposed side-information-based PLC. (PESQ = 3.00, PLCMOS = 4.11, STOI = 0.96).
    Compared with (b)--(d), (e) and (f) exhibit richer spectral details and higher perceptual scores, demonstrating the effectiveness of leveraging side information for in-band FEC.
    } 
	\label{fig:spectrogram}
\end{figure*}

Codec-specific in-band FEC improves robustness by embedding a compact representation of the previous frame within the current packet, which enables recovery from packet loss without retransmission. However, traditional codecs often struggle to accommodate such redundancy under strict bitrate budgets. Neural codecs, benefiting from superior compression efficiency, are well-suited to this paradigm. In our design, the side information $\bm{z}$, originally introduced for entropy modeling, is reused as in-band FEC, thereby enhancing resilience without introducing an additional codec.

As shown in Fig.~\ref{fig:side_info_learning}, side information $\bm{z}$ is jointly optimized through rate and distortion terms. RVQ is employed to constrain its bitrate to a low range. To improve robustness, random masking with a ratio uniformly sampled from 0 to 0.1 is applied during training to simulate partial packet loss of $\bm{z}$, considering its lower loss probability compared with the main stream. The loss function includes both rate loss and mean squared error (MSE) supervision. Since the latent $\bm{y}$ follows a Gaussian distribution under the variational or RD objective, the MSE term can be derived from the Gaussian-form KL divergence between the predicted and true distributions. This design increases accurate mean estimation and enhances the representational fidelity of $\bm{z}$, enabling it to serve as informative, low-bitrate redundancy for recovering corrupted content.

Fig.~\ref{fig:spectrogram} illustrates the perceptual advantages of the proposed design. Compared with zero-filling strategies applied to either the waveform or latent domains, the proposed method preserves finer harmonic structures and suppresses spectral artifacts. Both the spectrogram comparisons and objective metrics (PESQ, PLCMOS, and STOI) confirm that reusing $\bm{z}$ as in-band FEC significantly enhances reconstruction quality and plays a key role in improving error resilience for real-time speech transmission.

\begin{figure*}[t]
	\setlength{\abovecaptionskip}{0cm}
	\setlength{\belowcaptionskip}{0.cm}
	\hspace{-1em}
	\centering
	\includegraphics[width=0.9\linewidth]{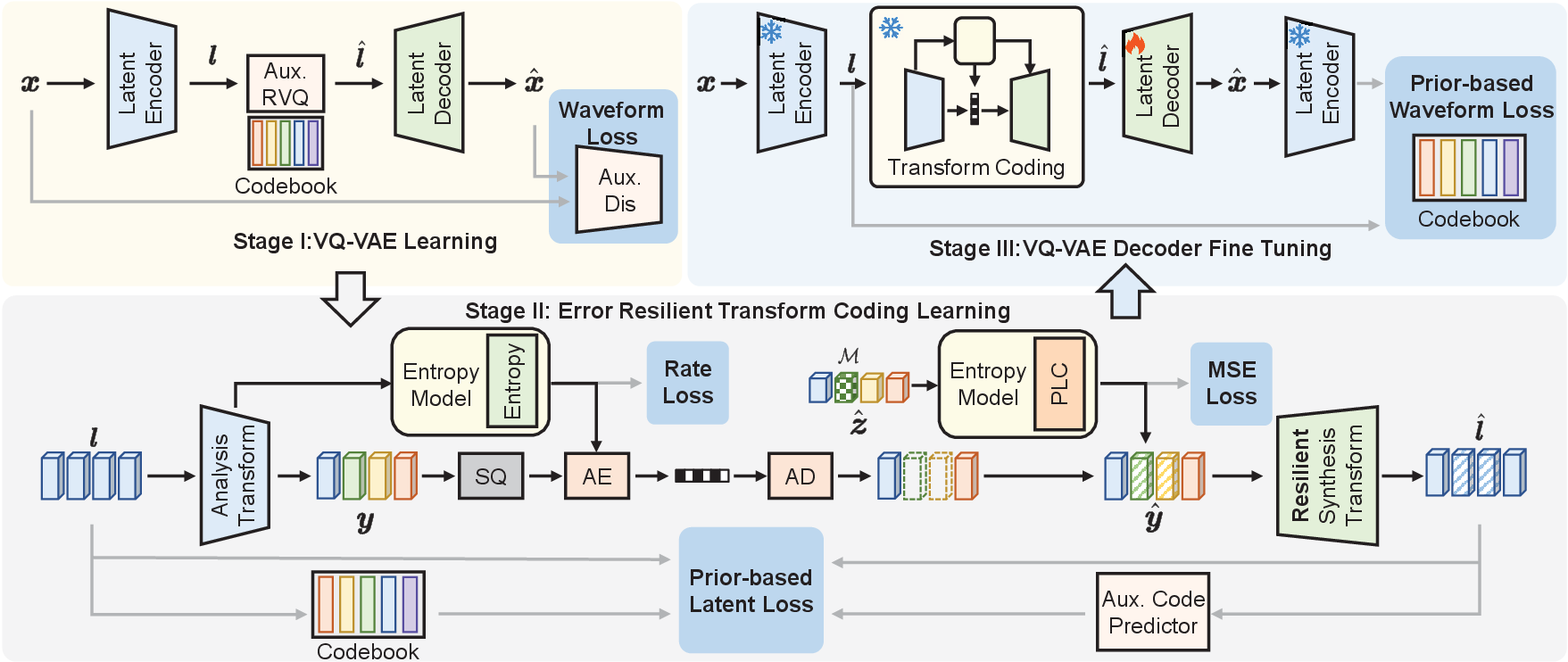}
	\caption{
	Progressive training pipeline consisting of three stages. 
    Stage I: The VQ-VAE is trained to learn generative latent representations from waveform data. 
    Stage II: The error-resilient transform coding is trained for latent compression and recovery of lost information, optimized under the RD objective, where the distortion term corresponds to the proposed prior-based latent loss. 
    Random mask embeddings are introduced to indicate missing latents and their confidence levels, and an additional MSE loss is applied to train the PLC module. 
    Stage III: The VQ-VAE decoder is fine-tuned with prior-based waveform supervision to align its reconstruction behavior with the distortion characteristics introduced by transform coding.
	}\label{fig:training_pipeline}
\end{figure*}

\subsection{Controllable Redundancy for Adaptive Error-Resiliency}\label{sec:Controllable Redundancy for Adaptive Error-Resiliency}

The redundancy level is determined by the bitrate of side information the number of side information copies $\{\bm{z}_{t-k} \mid k \in \mathcal{K}\}$ embedded in each frame $\bm{y}_t$, where $\mathcal{K} = \{k_1, k_2, \dots, k_N\}$ denotes a predefined set of frame offsets. In our configuration, the side information codebook size is fixed to $2^{10}$, corresponding to $0.5 \times Q$ kbps per $\bm{z}_t$ copy, where $Q$ is the number of RVQ quantizers affecting the reconstruction quality of lost frame $\bm{y}_t$. The total added redundancy is computed as
\begin{equation}
\text{Redundant Bitrate} = 0.5 \times Q \times N \ \text{kbps}.
\end{equation}

Assuming an i.i.d. packet-loss channel with a loss probability $p$, the probability that all $N$ redundant copies are lost is $p^N$, indicating that only a few copies are sufficient to ensure high reliability of $\bm{z}_t$. For burst-loss channels, the offset set $\mathcal{K}$ is chosen to be non-consecutive (e.g., $\mathcal{K} = \{1,13\}$ in our setup). This source-level control enables \emph{Glaris} to adapt protection strength under diverse channel conditions.

\subsection{Training Strategy} \label{sec:Training Strategy}
In this section, we detail a three-stage progressive training strategy to fully leverage the potential of generative latent space, as illustrated in Fig.~\ref{fig:training_pipeline}. 
Stage~I: A VQ-VAE is pre-trained to produce a generative latent representation.
Stage~II: Transform coding is learned via the latent alignment loss.
Stage~III: VQ-VAE decoder is fine tuned to align the mismatch caused by transform coding.

\subsubsection{Stage I: VQ-VAE Learning}
We train a generative VQ-VAE with adversarial learning at the bitrate $24$ kbps quantized by RVQ for high quality. The waveform loss comprises time-domain reconstruction loss $\ell_t$, frequency-domain reconstruction loss $\ell_f$, adversarial loss $\ell_g$, feature matching loss $\ell_{feat}$ and VQ commitment loss $\ell_w$, detailed in~\cite{defossez2023high}:

\begin{equation}
\begin{split}
\mathcal{D}_{\bm{x}}(\bm{x}, \hat{\bm{x}}) = & \lambda_t \cdot \ell_t(\boldsymbol{x}, \hat{\boldsymbol{x}}) + \lambda_f \cdot \ell_f(\boldsymbol{x}, \hat{\boldsymbol{x}}) + \lambda_g \cdot \ell_g(\hat{\boldsymbol{x}}) \\
& + \lambda_{feat} \cdot \ell_{feat}(\boldsymbol{x}, \hat{\boldsymbol{x}}) + \lambda_w \cdot \ell_w(w),
\end{split}
\end{equation}
where $\lambda_t, \lambda_f, \lambda_g, \lambda_{feat}$ and $\lambda_w$ are the scalar coefficients to balance between the terms. We also utilize the loss balancer proposed in~\cite{defossez2023high} to stabilize training with weights $\lambda_t = 0.1, \lambda_f = 1, \lambda_g = 3, \lambda_{feat} = 3$ and $\lambda_w = 1$. 

\subsubsection{Stage II: Error-Resilient Transform Coding Learning}
We learn the error-resilient transform coding while fixing the generative latent codec. To improve global sequence consistence of reconstructed $\hat{\bm{l}}$, a prior-based latent loss is introduced by
\begin{equation}
\mathcal{D}_\text{prior}(\bm{l}, \hat{\bm{l}}) = \beta \cdot \text{CE}(M_{\bm{l}}, \hat{M}_{\hat{\bm{l}}}) + ||\bm{l} - \hat{\bm{l}}||_2^2,
\end{equation}
where $\text{CE}$ denotes the cross-entropy loss and $\beta$ defaults to 0.5. By introducing an auxiliary code predictor $CP$, We encode $\bm{l}$ into VQ-indices by $M_{\bm{l}} = \text{RVQ}(\bm{l})$ and predict these indices by $\hat{M}_{\hat{\bm{l}}} = CP(\hat{\bm{l}})$. We introduce random masking to simulate packet loss and denote $\hat{\bm{l}}_{\text{rec}}$ as the reconstructed output without masking and $\hat{\bm{l}}_{\text{con}}$ as the concealed output with masking. The random mask ratio is uniformly sampled from 0 to 0.1 for $\bm{z}$ and from 0.05 to 0.7 for $\bm{y}$. Then the total distortion on $\bm{l}$ is:
\begin{equation}
\mathcal{L}_{\bm{l}} = \alpha \cdot \mathcal{D}_\text{prior}(\bm{l}, \hat{\bm{l}}_\text{con}) + \mathcal{D}_\text{prior}(\bm{l}, \hat{\bm{l}}_\text{rec}),
\end{equation}
where $\alpha$ is the scalar coefficient to balance compression and PLC performance. 
To train the PLC module, we also introduce MSE loss on $\bm{y}$, thus the total distortion is:
\begin{equation}
\mathcal{D}\ = \mathcal{L}_{\bm{l}} + \gamma \cdot ||\bm{y} - \hat{\bm{y}}||_2^2,
\end{equation}
where $\gamma$ is the scalar coefficient that defaults to 0.5.
The final RD trade-off is
\begin{equation}
\mathcal{L}_{\text{RD}} = \mathbb{E}_{x \sim p_X} \left[ \lambda \cdot \mathcal{R}(\bm{y}_Q) +  \mathcal{D}\right],
\end{equation}
where $\mathcal{R}$ is the rate loss and $\lambda$ is used to control the trade-off. We omit the codebook loss of $\bm{z}$ for the sake of conciseness.

\subsubsection{Stage III: VQ-VAE Decoder Fine Tuning}
We fine-tune only the latent decoder to achieve better performance. To leverage prior-based latent loss, we transfer $\bm{l}$ constraints into the source space. Specifically, we reuse the latent encoder $E$ to encode the generated $\hat{\bm{x}}$ into generative latent space, so that the $\bm{l}$ distortion can be calculated as $\mathcal{L}_{\bm{l}}$. Similarly, we denote $\hat{\bm{x}}_{\text{con}}$ and $\hat{\bm{x}}_{\text{rec}}$ for concealed and reconstructed $\hat{\bm{x}}$. The prior-based waveform loss that combines source and latents is:
\begin{align}
\mathcal{L}_{\bm{x}} = \alpha \cdot \mathcal{D}_{\bm{x}}(\bm{x},& \hat{\bm{x}}_\text{con}) + \mathcal{D}_{\bm{x}}(\bm{x}, \hat{\bm{x}}_\text{rec}),\\
\mathcal{L} = &\mathcal{L}_{\bm{x}}+\lambda_{\bm{l}}\cdot \mathcal{L}_{\bm{l}},
\end{align}
where $\lambda_{\bm{l}}$ defaults to 0.05.

\section{Experimental Results}\label{sec:exp}
\subsection{Experimental Settings} 

\subsubsection{Dataset and Training Details} 
The training dataset employed in this study comprises 360 hours of 16 kHz clean speech, extracted from the standard LibriSpeech dataset~\cite{panayotov2015librispeech}. LibriSpeech originates from the LibriVox project, which encompasses English audiobook recordings  contributed by online volunteers under copyright-free licenses. Specifically, the train-clean-100 and train-clean-360 subsets were utilized for training, while the test-clean subset was reserved for testing. 

The packet loss traces utilized in this study are derived from both simulated data and actual traces from the PLC challenge dataset provided for the Microsoft PLC challenge 2022~\cite{diener2022interspeech}. In terms of simulated loss traces, we incorporated memoryless i.i.d. packet-loss channels with a specified loss ratio as well as a three-state Markov  model~\cite{milner2004analysis}. To evaluate the performance under wireless channel conditions, we conduct experiments over the COST2100~\cite{liu2012cost} fading channel. CSI samples are collected in an indoor scenario at the 5.3\,GHz bands, and all schemes use a one-shot transmission. We simulate 5G link adaptation under the COST2100 channel using the official Sionna~\cite{hoydis2022sionna} library. The inner-loop adaptation selects the highest Modulation and Coding Scheme achieving a BLER below 0.1, based on the SNR feedback from the previous slot. Experiments are conducted at an average SNR of 8\,dB on a single-input single-output link. Each encoded frame is transmitted within one slot, and the resulting Hybrid Automatic Repeat Request trace is used as the packet loss trace.

Our model is trained with the Adam optimizer with a batch size of 8 examples of 2 seconds each, a learning rate of $3\cdot10^{-4}$, $\beta_{1} = 0.5$, and $\beta_{2} = 0.9$. For the discriminator and code prediction model, the learning rates are set to $10^{-4}$ and $5\cdot10^{-5}$, respectively. During variable rate training, $\lambda$ is sampled within the interval $[0.002,0.07]$, across 64 quantized levels.

\subsubsection{Metrics}
For speech quality assessment, we employ multiple metrics to provide a comprehensive performance evaluation:
\begin{itemize}
    \item PESQ: For perceptual quality, we use the Perceptual Evaluation of Speech Quality (PESQ) metric~\cite{rix2001perceptual}. PESQ, as defined in ITU-T P.862, evaluates speech quality in telephone systems and codecs, producing scores between 1 and 4.5 based on human auditory perception.
    \item STOI and WER: For intelligibility assessment, we use Short-Time Objective Intelligibility (STOI)~\cite{taal2010short} and Word Error Rate (WER). STOI evaluates intelligibility by correlating processed speech with reference speech, while WER is calculated based on a pretrained automatic speech recognition (ASR) model\footnote{\url{https://huggingface.co/jonatasgrosman/wav2vec2-large-xlsr-53-english}} fine-tuned on English from XLSR~\cite{conneau2020unsupervised}.
    \item MOS: For objective Mean Opinion Score (MOS), we use DNSMOS~\cite{reddy2022dnsmos}, NISQA~\cite{mittag2021nisqa}, and PLCMOS~\cite{diener2023plcmos}. DNSMOS predicts speech quality based on ITU-T P.808 standards. It has been upgraded to the P.835 standard, which specifies three distinct scores: speech quality (SIG), background noise quality, and overall audio quality (OVRL). NISQA assesses speech quality through an overall MOS score (OVRL), complemented by detailed evaluations of four specific dimensions: Noisiness (NOI), Coloration (COL), Discontinuity (DIS), and Loudness (LOU). PLCMOS focuses on MOS in packet loss scenarios.
    \item Subjective Evaluation: For subjective assessment, we conduct a MUSHRA test~\cite{series2014method}.
\end{itemize}

\subsubsection{Baseline Methods}
To establish a comprehensive benchmark for comparison, our experiments incorporate several baseline methods. The Opus codec~\cite{valin2012definition}, a state-of-the-art traditional speech codec widely adopted in VoIP applications, is included. Specifically, we use Opus 1.5, which has been enhanced with neural PLC FARGAN~\cite{valin2024very} and deep redundancy-based in-band FEC DRED~\cite{valin2024dred}. For the naive baseline, we employ SoundStream~\cite{zeghidour2021soundstream}, where lost latent vectors are simply replaced with zeros. 
SoundStream with entropy coding (EC), labeled SoundStream + EC, is also reported to have the best compression efficiency. 
Among pure neural PLC baselines, we select FD-PLC~\cite{xue2022towards} and SoundSpring~\cite{yao2025soundspring} for fair comparison, as both operate on SoundStream's latent space. FD-PLC can be viewed as a regression problem predicting vectors in the latent space while jointly training the decoder. SoundSpring, in contrast, is framed as a classification problem predicting indices and follows a plug-and-play approach. 
In addition, we include DeepSC-S~\cite{weng2021semantic}, a representative JSCC-based semantic communication system, serving as a reference for comparing non-streaming JSCC systems.

\begin{figure}[t]
	\setlength{\abovecaptionskip}{0cm}
	\setlength{\belowcaptionskip}{0.cm}
	\hspace{-1em}
	\centering
	\includegraphics[width=\linewidth]{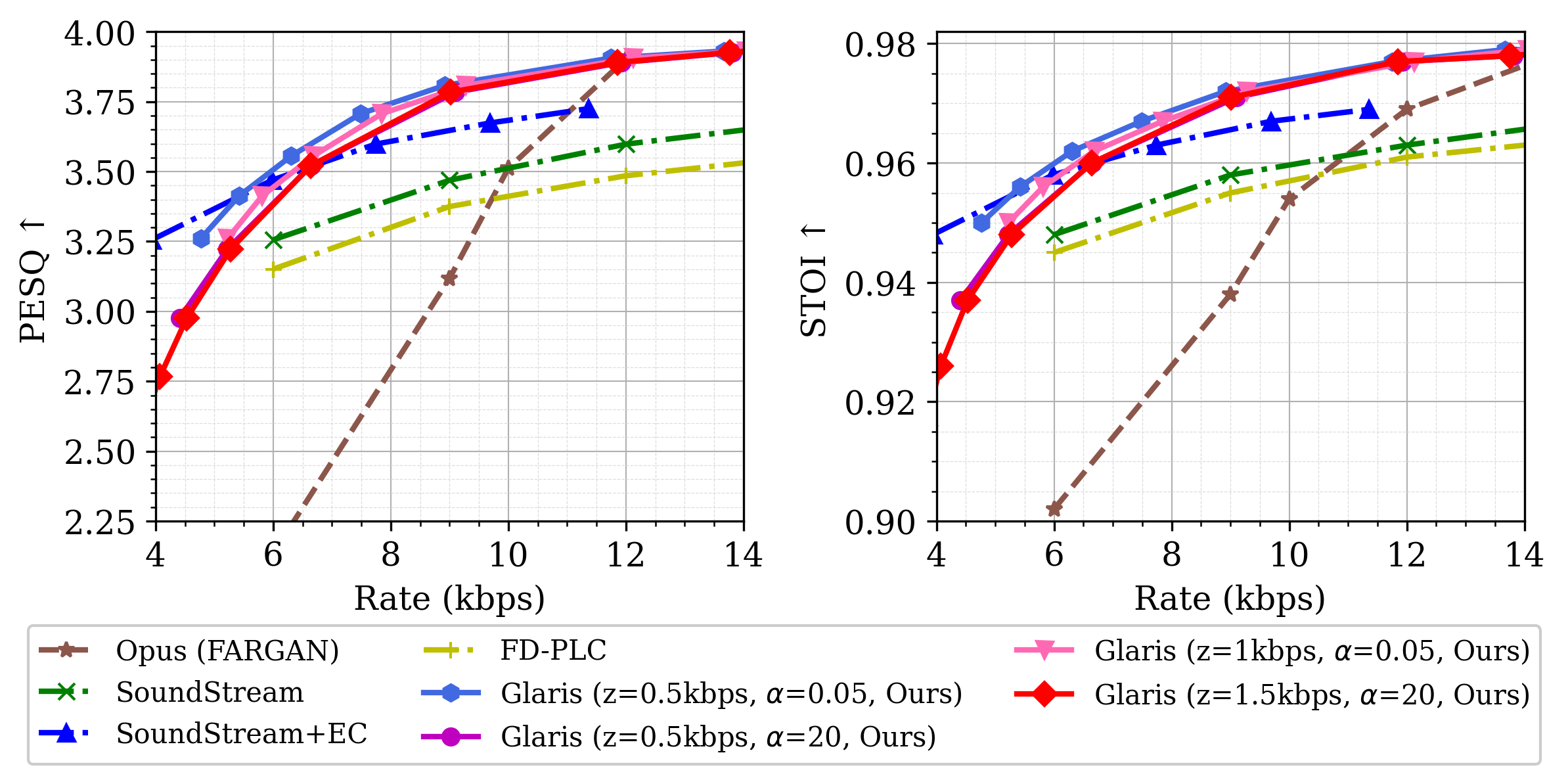}
	
	\caption{RD performance comparison in a reliable transmission scenario without packet loss. The proposed \emph{Glaris} with different settings consistently outperforms baselines, including Opus (FARGAN), SoundStream-based variants, and FD-PLC in terms of PESQ and STOI scores across various bitrates. Notably, the use of side information at different bitrates, in the absence of additional in-band FEC, does not degrade compression efficiency.}
	\label{fig:RD_PESQ_0}
\end{figure}
\subsection{Efficiency and Resilience Comparison} 

\subsubsection{Rate-Distortion Performance Comparison}
\begin{figure*}[t]
	\setlength{\abovecaptionskip}{0cm}
	\setlength{\belowcaptionskip}{0.cm}
	\hspace{-1em}
	\centering
	\includegraphics[width=\linewidth]{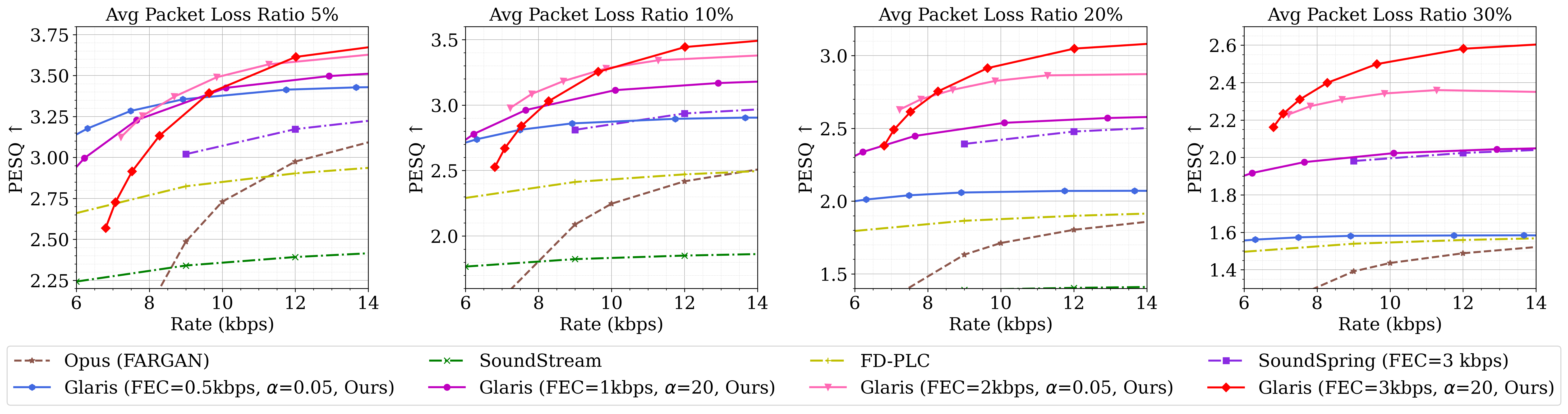}
	
	\caption{
	RD performance comparison under different average packet loss ratios. 
    The proposed \emph{Glaris} consistently outperforms baseline methods when the in-band FEC is properly configured. 
    As the packet loss ratio increases, the corresponding RD curves become progressively flatter, indicating that enhancing in-band FEC plays a more critical role than expanding the source bandwidth under lossy channel conditions.
	}
	\label{fig:RD_PESQ}
\end{figure*}
\begin{figure*}[t]
	\setlength{\abovecaptionskip}{0cm}
	\setlength{\belowcaptionskip}{0.cm}
	\hspace{-1em}
	\centering
	\includegraphics[width=\linewidth]{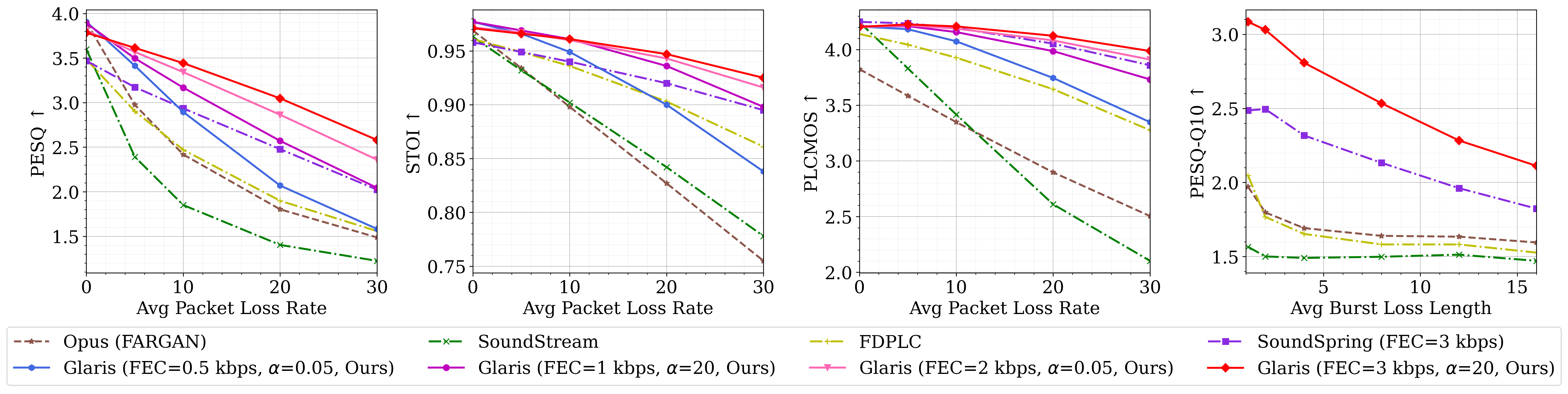}
	
	\caption{
	Objective quality comparison at 12 kbps under different packet loss rates and burst loss lengths. 
    The last subfigure is evaluated using a Markov packet-loss model with an average packet loss rate of 10\%. 
    Evaluation metrics include PESQ, STOI, PLCMOS, and PESQ-Q10, where PESQ-Q10 denotes the 10th percentile of PESQ scores across all speech segments, reflecting perceptual quality under burst loss conditions.
    }
	\label{fig:Decay_12kbps}
\end{figure*}
To illustrate the compression efficiency, the RD performance under error-free transmission is presented in Fig.~\ref{fig:RD_PESQ_0}. 
Except for Opus, all methods share the same SoundStream backbone. 
\emph{Glaris} consistently outperforms both traditional and neural codecs in compression efficiency. 
Error-resilient neural codecs such as FD-PLC typically improve robustness at the expense of reduced efficiency. 
In contrast, \emph{Glaris} maintains high compression efficiency while providing strong error resilience. 
In its most robust configuration, SoundSpring employs a language model for PLC in a plug-and-play manner without modifying the encoder or decoder, thereby achieving performance same to SoundStream. 
The SoundStream + EC achieves the highest compression efficiency among the baselines by applying entropy coding to the quantization indices, but this design also makes it highly vulnerable to packet loss. 
\emph{Glaris} further improves upon this through RD-optimized transform coding, surpassing SoundStream + EC and suppressing error propagation via side-information-based PLC and in-band FEC, thereby achieving both higher efficiency and stronger resilience.

The RD performance under different average packet loss ratios is shown in Fig.~\ref{fig:RD_PESQ}. 
\emph{Glaris} consistently achieves higher PESQ scores than existing baselines, demonstrating superior robustness under unreliable transmission. 
As the packet loss ratio increases, the marginal benefit of increasing the source rate diminishes, particularly for baseline methods, whereas \emph{Glaris} maintains relatively high perceptual quality even at moderate bitrates. 
This trend underscores that error resilience plays a more critical role than compression efficiency in lossy environments.

We further demonstrate that the explicit in-band FEC redundancy in \emph{Glaris} is more effective than the implicit redundancy learned by the neural encoder. 
The neural encoder learns to inject redundancy into $\bm{y}$ by introducing packet loss during training and optimizing a corresponding loss function. 
However, as shown in Fig.~\ref{fig:RD_PESQ}, \emph{Glaris} variants indicate that increasing the FEC bitrate (e.g., 2~kbps vs. 1~kbps) yields substantial robustness gains even when the error-resilient regularization strength $\alpha$ is reduced. 
The limitation of implicit redundancy stems from the causal and limited-context nature of streaming sequence modeling, as well as from the intrinsic challenge of jointly optimizing compression efficiency and error resilience during training.

While \emph{Glaris} demonstrates strong robustness, excessive in-band FEC can degrade quality in low-bitrate scenarios due to bandwidth overhead from redundancy.
This suggests the need for adaptive in-band FEC control rather than fixed absolute FEC settings to balance source fidelity and resilience. 
Nevertheless, \emph{Glaris} exhibits higher tolerance to FEC configuration than channel-level FEC methods, providing flexible and robust adaptation across diverse network conditions.

\subsubsection{Robustness to Transmission Errors}

\paragraph{Error Resilience under Variable Loss Rate and Burst Loss Length}
To further assess the error-resilience characteristics of the proposed framework, reconstructed speech quality is evaluated at 12~kbps under varying packet loss rates, using three objective metrics: PESQ, STOI, and PLCMOS, which quantify perceptual quality, intelligibility, and subjective perceptual quality, respectively.

As shown in Fig.~\ref{fig:Decay_12kbps}, SoundStream with zero-filling performs well in the lossless case but degrades rapidly as the loss rate increases, revealing its sensitivity to channel impairments in the absence of dedicated resilience mechanisms. In contrast, codecs incorporating neural PLC, such as FD-PLC and Opus (FARGAN), exhibit smoother performance decay and higher robustness.

Among the baselines, FD-PLC achieves consistently higher scores than Opus, validating the benefit of end-to-end optimization. 
\emph{Glaris} is evaluated under multiple in-band FEC configurations to analyze its scalability. With a 0.5~kbps FEC budget, \emph{Glaris} attains performance comparable to FD-PLC while retaining higher compression efficiency in the lossless setting. Increasing the FEC rate to 1~kbps achieves comparable performance to SoundSpring, which uses 3~kbps redundancy. The 3~kbps configuration of \emph{Glaris} surpasses all baselines across severe loss conditions, confirming its scalable robustness with increased side-information-based in-band FEC.

To examine resilience under burst losses, PESQ-Q10 is measured using a Markov packet-loss model with a fixed average loss rate of 10\%, where PESQ-Q10 denotes the 10th-percentile PESQ across all speech segments, reflecting quality under long-burst degradations. The results in Fig.~\ref{fig:Decay_12kbps} show that \emph{Glaris} consistently maintains the highest PESQ-Q10 values, demonstrating the effectiveness of the proposed in-band FEC in mitigating burst loss impact.



\begin{table}[t]
	\scriptsize
	\centering
	\caption{
		DNSMOS Results under Different Loss Rates at 12 kbps.
	}
	\tabcolsep=0.08cm
	\resizebox{\linewidth}{!}{
		\begin{tabular}{m{1.8cm}<{\centering}m{0.6cm}<{\centering}m{0.7cm}<{\centering}m{0.7cm}<{\centering}m{0.7cm}<{\centering}m{0.7cm}<{\centering}m{0.7cm}<{\centering}m{0.7cm}}
			\toprule
			
			 & & \multicolumn{2}{c}{P.808 MOS$\uparrow$} & \multicolumn{2}{c}{SIG$\uparrow$} & \multicolumn{2}{c}{OVRL$\uparrow$}  \\
			\cmidrule(lr){3-4}
			\cmidrule(lr){5-6}
			\cmidrule(lr){7-8}
			{Method} & {FEC} & {5\%} & {30\%}  & {5\%}  & {30\%} & {5\%}  & {30\%} \\
			\midrule
			Opus (FARGAN) & / & 3.64 & 3.38 & 3.52  & 3.33 & 3.18 & 2.89 \\
			
			SoundStream & / & 3.76 & 3.28 & 3.53 & 2.96 & 3.16 & 2.3 \\
			
			FD-PLC & / & 3.81 & 3.67 & 3.6 & 3.45 & 3.3 & 3.07 \\
			
			SoundSpring &25\%  & \textbf{3.87} & 3.76 & 3.6 & 3.5 & 3.31 & 3.15 \\
			\midrule
			
			\emph{Glaris} &4.2\% & \underline{3.85} & 3.69 & \textbf{3.61} & 3.48 & \textbf{3.31} & 3.13 \\
			  &8.3\% & 3.85 & 3.77 & \underline{3.61} & 3.53 & \underline{3.31} & 3.2 \\
			  &16.7\% & 3.85 & \underline{3.79} & 3.61 & \underline{3.56} & 3.31 & \underline{3.24} \\
			  &25\% & 3.85 & \textbf{3.8} & 3.61 & \textbf{3.57} & 3.31 & \textbf{3.26} \\
			\bottomrule
		\end{tabular}
	}
	\label{tab:DNSMOS}
\end{table}
\begin{table}[t]
	\scriptsize
	\centering
	\caption{
		NISQA Results Under 30\% Loss Rates at 12 kbps.
	}
	\tabcolsep=0.08cm
	\resizebox{\linewidth}{!}{
		\begin{tabular}{m{1.8cm}<{\centering}m{0.6cm}<{\centering}m{0.84cm}<{\centering}m{0.84cm}<{\centering}m{0.84cm}<{\centering}m{0.84cm}<{\centering}m{0.96cm}}
			\toprule
			
			{Method} & {FEC} & {OVRL$\uparrow$} & {NOI$\uparrow$} & {COL$\uparrow$}  & {DIS$\uparrow$} & {LOU$\uparrow$} \\
			\midrule
			Opus (FARGAN) & / & 1.77 & 2.88 & 1.91 & 2.07 & 3.13 \\
			
			SoundStream & / & 2.18 & 2.74 & 2.59 & 2.58 & 3.34 \\
			
			FD-PLC & / & 3.43 & 3.37 & 3.71 & 3.31 & 3.94 \\
			
			SoundSpring &25\%  & 3.98 & \textbf{3.76} & 4.14 & 3.74 & 4.2\\
			\midrule
			
			\emph{Glaris} &4.2\% & 3.48 & 3.42 & 3.8 & 3.31 & 3.94  \\
			 &8.3\% & 3.89 & 3.65 & 4.16 & 3.68 & 4.16 \\
			 &16.7\% & \underline{4.04} & 3.72 & \underline{4.28} & \underline{3.81} & \underline{4.23} \\
			 &25\% & \textbf{4.09} & \underline{3.75} & \textbf{4.32} & \textbf{3.87} & \textbf{4.26} \\
			\bottomrule
		\end{tabular}
	}
	\label{tab:NISQA}
\end{table}
\begin{figure}[t]
	\setlength{\abovecaptionskip}{0cm}
	\setlength{\belowcaptionskip}{0.cm}
	\hspace{-1em}
	\centering
	\includegraphics[width=0.95\linewidth]{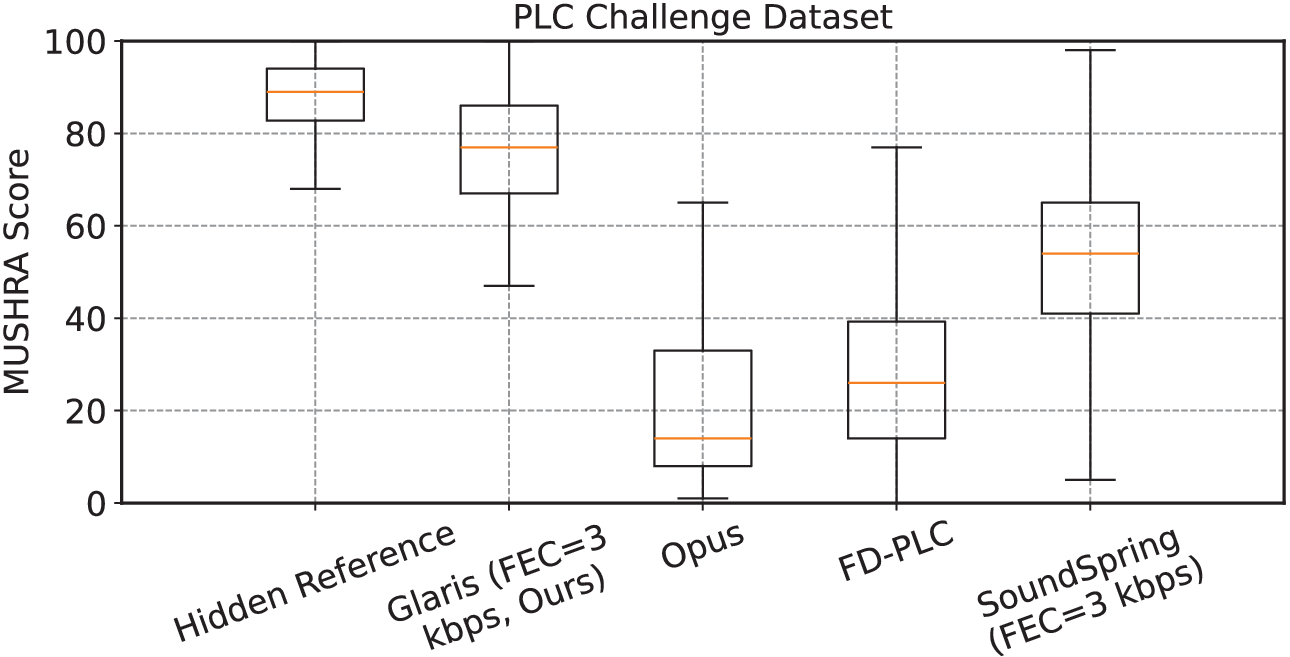}
	
	\caption{Result of subjective listening tests at 12 kbps. Demo examples of the reconstructed speech are available for comparison at \url{https://semcomm.github.io/Glaris}.}
	\label{fig:MUSHRA}
\end{figure}



\paragraph{MOS Evaluation}
To assess perceptual quality from both objective and subjective perspectives, DNSMOS and NISQA results at 12~kbps are presented in Tables~\ref{tab:DNSMOS} and~\ref{tab:NISQA}, and subjective MUSHRA scores are shown in Fig.~\ref{fig:MUSHRA}.

From Table~\ref{tab:DNSMOS}, methods incorporating FEC generally achieve higher perceptual scores. 
\emph{Glaris} maintains strong performance across both low and high loss rates, with its advantage becoming more pronounced at 30\% loss. 
Notably, it achieves comparable or superior MOS results with a lower FEC cost than SoundSpring, demonstrating a more efficient redundancy design.

Table~\ref{tab:NISQA} further evaluates perceptual dimensions under 30\% packet loss. 
\emph{Glaris} attains the highest overall NISQA score and consistently outperforms all baselines in coloration, discontinuity, and loudness, while maintaining competitive noisiness performance. 
These findings indicate that \emph{Glaris} effectively preserves perceptual fidelity under severe loss conditions with substantially lower redundancy overhead.

Subjective results in Fig.~\ref{fig:MUSHRA}, evaluated on actual packet-loss traces from the PLC Challenge dataset, further confirm these observations. 
\emph{Glaris} achieves the highest MUSHRA score among all baselines, approaching the perceptual quality of the hidden reference and validating its robustness and perceptual consistency under realistic transmission conditions.

\paragraph{Efficiency of In-band FEC}
To evaluate the efficiency of the proposed side-information-based in-band FEC, different FEC configurations are compared at a fixed total bitrate of 18~kbps, as shown in Table~\ref{tab:DRED}. 
For fairness, PESQ is adopted as the evaluation metric since it favors traditional waveform codecs such as Opus. 
To enable DRED in Opus, its bitrate is set to 19.5~kbps, while all other methods are constrained to 18~kbps. 
Across all settings, the number of redundant packets is kept the same to ensure comparability.

The results indicate that \emph{Glaris} with 1~kbps FEC achieves performance comparable to Opus (DRED) and SoundSpring. 
As the in-band FEC bitrate increases, \emph{Glaris} consistently surpasses both methods, demonstrating the high efficiency of its learned redundancy. 
This improvement suggests that \emph{Glaris} learns a more effective redundancy than the coarse layers of RVQ tokens used in SoundSpring, where fine-grained RVQ layers are difficult to predict based on coarse RVQ layers. 
Compared with Opus (DRED), \emph{Glaris} benefits from end-to-end optimization and the use of generative latent priors, enabling higher perceptual quality with less redundancy, validating the efficiency of the proposed in-band FEC design.

\begin{table}[t]
	\scriptsize
	\centering
	\caption{
		PESQ Results under Different Loss Rates at 18 kbps. 
	}
	\tabcolsep=0.08cm
	\resizebox{0.87\linewidth}{!}{
	    \begin{threeparttable}[b]
		\begin{tabular}{m{1.8cm}<{\centering}m{0.6cm}<{\centering}m{0.84cm}<{\centering}m{0.84cm}<{\centering}m{0.84cm}<{\centering}m{0.84cm}}
			\toprule
			
			{Method} & {FEC}\tnote{a} & {5\%} & {10\%}  & {20\%} & {30\%} \\
			\midrule
			Opus (DRED) & 5.85  & 3.41 & 2.99 & 2.48 & 2.14 \\
			
			SoundSpring   &3 & 3.3 & 3.01 & 2.53 & 2.06 \\
			\midrule
			
			\emph{Glaris} &1  & 3.54 & 3.2 & 2.6 &  2.07\\
			 &2 & \underline{3.65}& \underline{3.39} & \underline{2.86} & \underline{2.33} \\
			 &3 & \textbf{3.74} & \textbf{3.54} & \textbf{3.12} & \textbf{2.63} \\
			\bottomrule

		\end{tabular}
		\begin{tablenotes}
            \item[a] The FEC column indicates the in-band FEC in kbps.
        \end{tablenotes}
        \end{threeparttable}
	}
	\label{tab:DRED}
\end{table}

\begin{table}[t]
	\scriptsize
	\centering
	\caption{
		WER Results under Different Loss Rates at 12 kbps.
	}
	\tabcolsep=0.08cm
	\resizebox{\linewidth}{!}{
		\begin{tabular}{m{1.8cm}<{\centering}m{0.6cm}<{\centering}m{0.84cm}<{\centering}m{0.84cm}<{\centering}m{0.84cm}<{\centering}m{0.84cm}<{\centering}m{0.84cm}}
			\toprule
			
			{Method} & {FEC} & {0\%} & {5\%} & {10\%}  & {20\%} & {30\%} \\
			\midrule
			Opus (FARGAN) & / & 6.9\% & 7.6\% & 8.5\% & 11.4\% & 17.1\% \\
			
			SoundStream & / & 7\% & 7.6\% & 8.5\% & 11.1\% & 17.6\% \\
			
			FD-PLC & / & 6.9\% & 7.2\% & 7.6\% & 9.3\% & 12.3\% \\
			
			SoundSpring &25\%  & 7.2\% & 7.4\% & 7.7\% & 8.7\% & \underline{10.8\%} \\
			\midrule
			
			\emph{Glaris} &4.2\% & \textbf{6.7\%} & 7.3\% & 9\% & 15.7\% & 31.3\% \\
			 &8.3\% & \underline{6.8\%} & \textbf{6.9\%} & 7.4\% & 9.3\% & 15.4\% \\
			 &16.7\% & 6.8\% & \underline{6.9\%} & \underline{7.2\%} & \underline{8.3\%} & 12.2\% \\
			 &25\% & 6.8\% & 6.9\% & \textbf{7.1\%} & \textbf{7.9\%} & \textbf{10.5\%} \\
			\bottomrule
		\end{tabular}
	}
	\label{tab:WER}
\end{table}

\paragraph{Intelligibility Assessment}
To further evaluate speech intelligibility, we report the WER under different packet loss rates at 12~kbps, as shown in Table~\ref{tab:WER}. 
Since these models are not trained for ASR, the WER from a pretrained ASR model serves as an indicator of how well linguistic content is preserved in the reconstructed speech.

As presented in the table, \emph{Glaris} consistently achieves lower WER than baseline methods across most loss conditions, demonstrating its effectiveness in maintaining intelligible content. 
Incorporating in-band FEC generally improves performance. However, at high packet loss rates, when the inserted redundancy is insufficient, audible artifacts and distortions may occur.
This effect is particularly evident under a 30\% packet loss rate, where the 8.3\% FEC configuration results in a higher WER than FD-PLC because of insufficient side information for accurate recovery.

This limitation can be alleviated by increasing the bitrate of~$\bm{z}$ without requiring additional backup frames. 
When the fidelity of~$\bm{z}$ improves, masked-token prediction becomes more accurate and fewer recognition errors occur. 
This trend is confirmed in the 25\% FEC configuration, where \emph{Glaris} achieves the lowest WER under 30\% loss. 
These findings highlight the critical role of side-information bitrate in enhancing speech intelligibility under severe packet-loss conditions.




\begin{figure}[t]
	\setlength{\abovecaptionskip}{0cm}
	\setlength{\belowcaptionskip}{0.cm}
	\hspace{-1em}
	\centering
	\includegraphics[width=\linewidth]{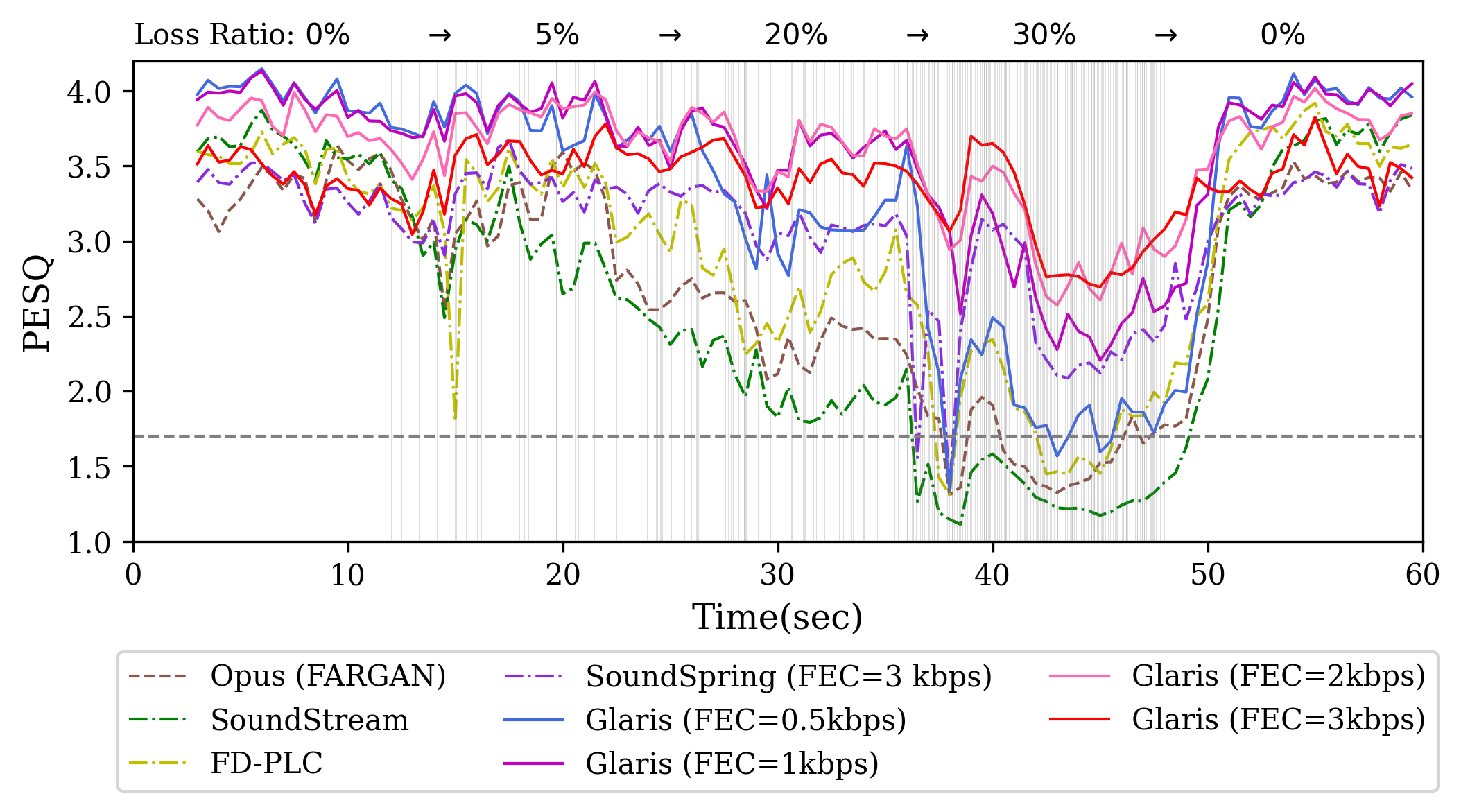}
	\caption{
	Real-time PESQ evaluation under a dynamic packet loss trace at 8 kbps. The loss pattern varies over time with labeled average loss ratios, and packet drop events are indicated by gray vertical lines. At each time step, the PESQ score is computed based on the latest 3-second audio segment. 
	}
	\label{fig:variable_loss_rate}
\end{figure}

\begin{figure}[t]
	\setlength{\abovecaptionskip}{0cm}
	\setlength{\belowcaptionskip}{0.cm}
	\hspace{-1em}
	\centering
	\includegraphics[width=\linewidth]{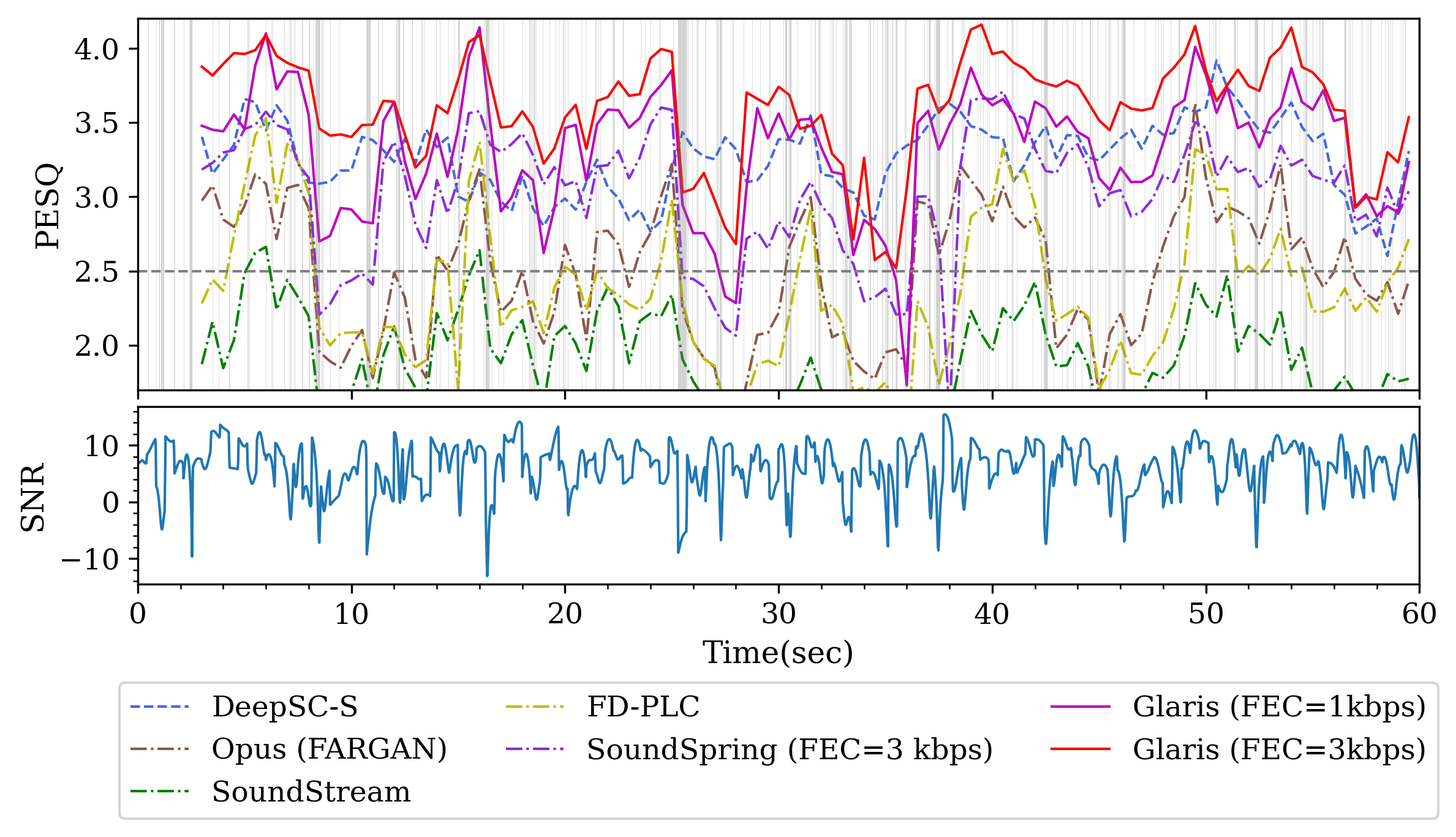}
	
	\caption{
	Real-time PESQ evaluation under a dynamic COST2100 wireless channel. The packet loss trace follows standard 5G link adaptation under time-varying channel conditions, and packet drop events are indicated by gray vertical lines.
	DeepSC-S, representing a JSCC-based semantic communication system, operates a bandwidth of 24 kHz, while all separation-based schemes use a 6 kHz bandwidth. PESQ scores are computed every 3 s using the latest audio segment, with loss events indicated by gray vertical lines.
	}
	\label{fig:cost2100_jscc}
\end{figure}

\begin{figure}[t]
	\setlength{\abovecaptionskip}{0cm}
	\setlength{\belowcaptionskip}{0.cm}
	\hspace{-1em}
	\centering
	\includegraphics[width=0.95\linewidth]{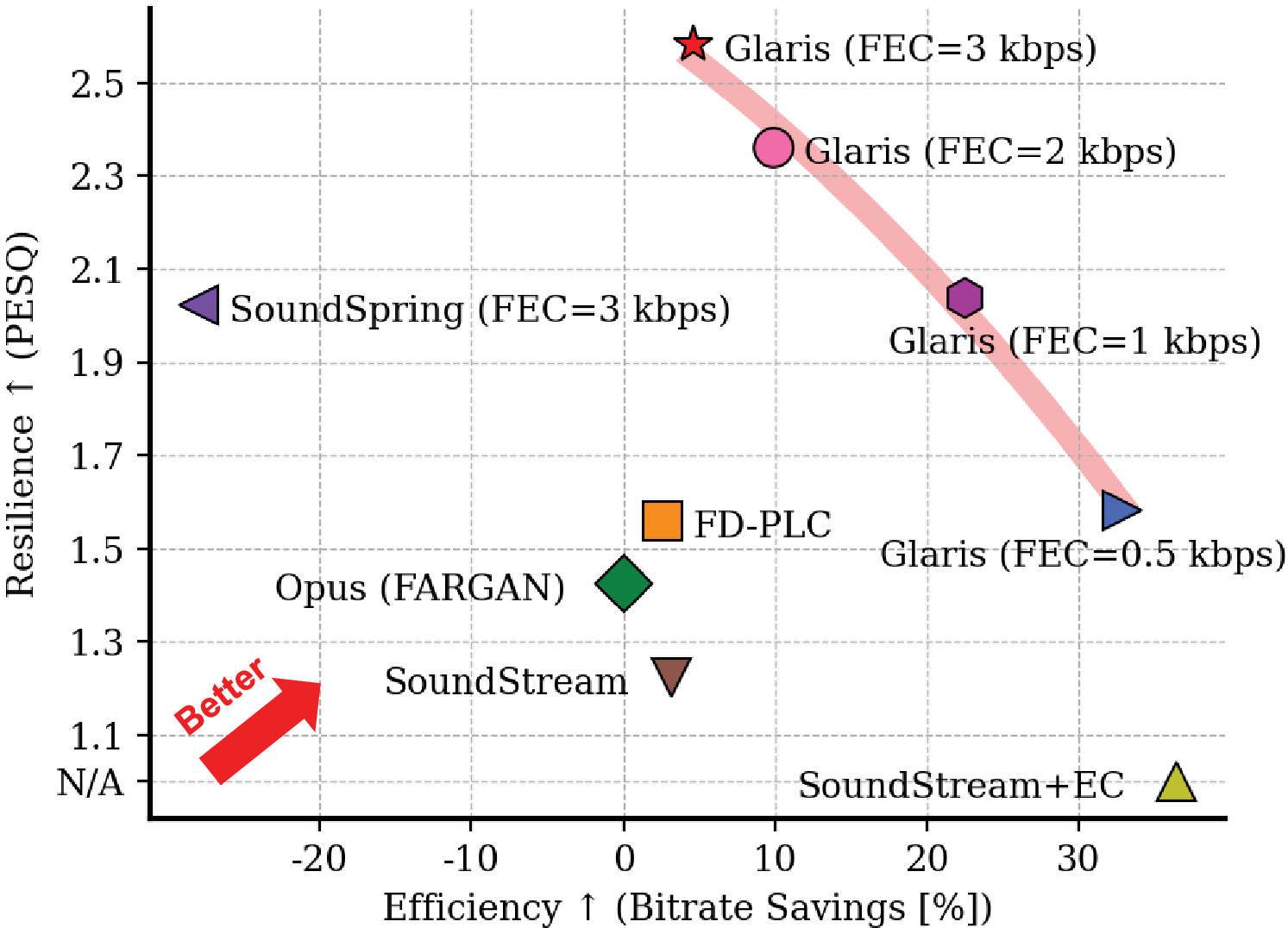}
	
	\caption{Efficiency-resilience trade-off comparison across different methods. Top-right is better. The proposed \emph{Glaris} framework achieves a favorable balance between efficiency and robustness by adjusting the amount of in-band FEC, offering flexible adaptation to different application requirements and network conditions.}
	\label{fig:benchmark}
\end{figure}
\paragraph{Real-time PESQ Evaluation}
As shown in Figs.~\ref{fig:variable_loss_rate} and~\ref{fig:cost2100_jscc}, \emph{Glaris} exhibits strong error resilience in both random packet-loss and time-varying wireless channel conditions. 
In the dynamic loss scenario, where the packet loss rate increases from 0\% to 30\% and then decreases, schemes without in-band FEC experience pronounced PESQ degradation and high sensitivity to loss variations, which result in audible interruptions. 
SoundSpring maintains relatively stable quality through its plug-and-play PLC mechanism, while \emph{Glaris} achieves comparable or even better performance with lower FEC overhead and superior quality during loss-free intervals. 

Under the practical COST2100 wireless channel, \emph{Glaris} sustains perceptual quality comparable to the JSCC-based DeepSC-S, despite operating with only one-fourth of its bandwidth.
These observations confirm that \emph{Glaris} effectively enhances error resilience while preserving compression efficiency by leveraging generative latent priors, thereby achieving JSCC-level robustness within a source-channel separated framework.


\subsubsection{Balancing Error Resilience and Efficiency}


To evaluate the trade-off between error resilience and compression efficiency, we present a comparison of various methods in Fig.~\ref{fig:benchmark}. Efficiency is measured by BD-rate savings relative to Opus, while resilience is quantified by PESQ scores at 12~kbps under a 30\% packet loss rate. The results demonstrate that \emph{Glaris} not only achieves significantly higher robustness compared to baseline methods, but also maintains a more favorable efficiency-resilience balance, outperforming approaches such as SoundSpring. Furthermore, this trade-off can be flexibly adjusted by varying the in-band FEC bitrate, enabling \emph{Glaris} to flexibly balance efficiency and robustness according to application requirements and channel conditions.

\subsection{Ablation Study}  
\subsubsection{Impact of Prior-based Latent Loss}
\begin{table}[t]
	\scriptsize
	\centering
	\caption{
		Ablation Study on Prior-Based Latent Loss.
	}
	\tabcolsep=0.08cm
	\resizebox{0.788\linewidth}{!}{
    	\begin{threeparttable}[b]
    		\begin{tabular}{m{1.3cm}<{\centering}m{1.3cm}<{\centering}m{1.3cm}<{\centering}m{1.3cm}<{\centering}}
    			\toprule
    			{$\bm{l}$ MSE} & {$\bm{l}$ CE loss} & {Efficiency\tnote{a} $\downarrow$} & {Resilience\tnote{b} $\uparrow$} \\
    			\midrule
    			\xmark & \xmark  & 35.6\% & 2.47\\
    			\cmark & \xmark & 28.6\% & 2.35\\
    			\cmark & \cmark & 0 & 2.58 \\
    			\bottomrule
    		\end{tabular}
		\begin{tablenotes}
          \item[a] Efficiency is evaluated in BD-rate.
          \item[b] Resilience is evaluated in in PESQ under 30\% packet loss at 12 kbps.
        \end{tablenotes}
    	\end{threeparttable}
	}
	\label{tab:ablation_latent_loss}
\end{table}  

Table~\ref{tab:ablation_latent_loss} summarizes the ablation study on the prior-based latent loss, where the contributions of the $\bm{l}$-space MSE and CE terms are evaluated separately. 
Removing both terms leads to the highest BD-rate, indicating inefficient compression. 
Using only the MSE term improves compression efficiency but degrades resilience, since it constrains the reconstruction to the mean of the Gaussian distribution and ignores global sequence modeling. 
Introducing the CE term aligns the predicted latent features with the prior distribution, which enhances perceptual quality and strengthens error resilience by improving latent consistency. 
When both terms are jointly applied, the model achieves the best overall trade-off, reducing the BD-rate by up to 35.6\% and improving PESQ under packet loss. 
These results confirm that the proposed prior-based latent loss is essential for jointly optimizing compression efficiency and robustness.



\subsubsection{Latency Analysis}
\begin{table}[t]
	\scriptsize
	\centering
	\caption{
		Real-Time Factor (RTF) for 20 ms Frames at 12 kbps in Streaming Inference.
	}
	\tabcolsep=0.08cm
	\resizebox{0.91\linewidth}{!}{
	    
    		\begin{tabular}{m{2cm}<{\centering}m{1.0cm}<{\centering}m{1.7cm}<{\centering}m{1.6cm}<{\centering}}
    			\toprule
    			 {Method} & {Enc.} & {Dec. (w/o PLC)} & {Dec. (w PLC)} \\
    
    			\midrule
    			SoundStream & 2.1 & 2.17 & /\\
    			SoundStream + EC & 1.15 & 1.06 & / \\
    			\midrule
    			\emph{Glaris} & 1.17 & 1.27 & 1.52 \\
    			\bottomrule
    		\end{tabular}
    }
%
    
	\label{tab:latency}
\end{table}  

To evaluate system latency, we report the RTF of different methods in Table~\ref{tab:latency}. 
RTF is defined as the ratio between input duration and processing time, where values above one indicate real-time capability. 
Because encoding and decoding run in parallel, overall latency is determined by the slower process. 
All measurements are performed on a four-thread Intel(R) Xeon(R) Gold~6226R CPU in a frame-by-frame inference manner.

As shown in Table~\ref{tab:latency}, \emph{Glaris} achieves real-time inference with an RTF comparable to SoundStream~+~EC. 
A slightly higher RTF is observed during decoding with PLC, as entropy decoding can be bypassed when packet-loss detection predicts failure, introducing minor computational overhead. 
These results demonstrate that \emph{Glaris} maintains real-time performance under causal inference and is well-suited for deployment in practical speech communication systems.

\section{Conclusion}\label{sec:con}
This paper presented \emph{Glaris}, an error-resilient neural speech communication framework that leverages generative latent priors to achieve a favorable balance between compression efficiency and transmission robustness. By jointly modeling the latent prior and hyperprior within a two-stage coding framework, \emph{Glaris} enhances semantic consistency and reconstruction fidelity under packet loss. The proposed side-information-based error resilience mechanism enables PLC and in-band FEC to work in concert, providing integrated sender-receiver protection, while the controllable redundancy mechanism allows for adaptive error resilience under diverse network conditions. Extensive experiments on the LibriSpeech dataset under multiple channel models, including both network and wireless channel models such as COST2100, demonstrate that \emph{Glaris} consistently outperforms existing codecs in both objective and subjective evaluations, achieving comparable robustness to JSCC in the separation-based method. In future work, we plan to extend \emph{Glaris} toward multi-modal and cross-lingual speech communication, and further explore scheduling algorithms that exploit enhanced error resilience for improved system-level performance.

\ifCLASSOPTIONcaptionsoff
\newpage
\fi
\bibliographystyle{IEEEtran}
\bibliography{Ref}


\end{document}